\documentclass[preprint,prd,nofootinbib,tightenlines,groupedaddress,amsmath,amssymb]{revtex4}

\usepackage{bm}
\usepackage{color}
\usepackage{graphicx}
\usepackage[hypertex]{hyperref}
\usepackage{multirow}

\begin{document}
\baselineskip=15pt \parskip=3pt

\vspace*{3em}

\title{Minimal Lepton Flavor Violation Implications\\of the \boldmath$b\to s$ Anomalies}

\author{Chao-Jung Lee and Jusak Tandean}
\affiliation{Department of Physics and Center for Theoretical  Sciences,
National Taiwan University, \\ Taipei 106, Taiwan \\
$\vphantom{\bigg|_{\bigg|}^|}$}

\begin{abstract}
The latest measurements of rare \,$b\to s$\, decays in the LHCb experiment have led to results
in tension with the predictions of the standard model (SM), including a tentative indication
of the violation of lepton flavor universality.
Assuming that this situation will persist because of new physics, we explore some of
the potential consequences in the context of the SM extended with the seesaw mechanism involving
right-handed neutrinos plus effective dimension-six lepton-quark operators under the framework
of minimal flavor violation.
We focus on a couple of such operators which can accommodate the LHCb anomalies and conform to
the minimal flavor violation hypothesis in both their lepton and quark parts.
We examine specifically the lepton-flavor-violating decays \,$B\to K^{(*)}\ell\ell'$,
\,$B_s\to\phi\ell\ell'$, \,$B\to(\pi,\rho)\ell\ell'$,\, and \,$B_{d,s}\to\ell\ell'$,\,
as well as \,$K_L\to e\mu$\, and\,\,\,$K\to\pi e\mu$,\, induced by such operators.
The estimated branching fractions of some of these decay modes with $\mu\tau$ in the final
states are allowed by the pertinent experimental constraints to reach a few times\,\,$10^{-7}$
if other operators do not yield competitive effects.
We also look at the implications for \,$B\to K^{(*)}\nu\nu$\, and \,$K\to\pi\nu\nu$, finding
that their rates can be a few times larger than their SM values.
These results are testable in future experiments.
\end{abstract}

\maketitle

\section{Introduction\label{intro}}

The recently acquired data on a number of observables in \,$b\to s$\, decays have revealed
some intriguing tensions with the expectations of the standard model (SM).
Specifically, last year the LHCb Collaboration\,\,\cite{Aaij:2014ora} determined the ratio
of branching fractions of the decays \,$B^+\to K^+\mu^+\mu^-$\, and \,$B^+\to K^+e^+e^-$\,
to be \,$R_K^{}=0.745^{+0.090}_{-0.074}{\rm(stat)}\pm0.036{\rm(syst)}$\,
for the dilepton invariant mass squared range of 1-6$\rm\;GeV^2$.
This result diverges from the lepton universality in the SM by 2.6$\sigma$.
Moreover, earlier LHCb\,\,\cite{Aaij:2013qta} reported a local discrepancy at the 3.7$\sigma$
level from the SM prediction for one of the angular observables in the decay
\,$B^0\to K^{*0}\mu^+\mu^-$.\,
This disagreement has persisted after an updated analysis was done using the full LHCb
Run\,I dataset\,\,\cite{LHCb:2015dla}.
In addition, the latest measurements by LHCb\,\,\cite{Aaij:2013aln} of the branching fractions
of several rare \,$b\to s$\, decays favor values less than those estimated in the SM.

Although the statistical significance of these anomalies is still too low for a definite
conclusion, they may be hinting at the presence of physics beyond the SM.
Subsequent model-independent theoretical works have in fact shown that
new physics (NP) could resolve the tensions\,\,\cite{Descotes-Genon:2013wba,
Hiller:2014yaa,Altmannshofer:2014rta,Altmannshofer:2015sma,rsmodel}.\footnote{The tensions may
also be alleviated by including the effects of charm-anticharm resonances \cite{Lyon}.\smallskip}
In particular, NP contributing via dimension-six operators
of the form\,\,\cite{Hiller:2014yaa,Altmannshofer:2014rta,Altmannshofer:2015sma}
\begin{eqnarray} \label{Lsmnp}
{\mathcal L}_{\rm SM+NP}^{} \,\,=\,\, \frac{\alpha_{\rm e}^{}G_{\rm F}^{}V_{ts}^*V_{tb}^{}}
{\sqrt2\,\pi}\, \overline{s}_{\,}\gamma^\beta P_L^{} b\;\overline{\ell}_{\,}\gamma_\beta^{}
\big(C_9^\ell+C_{10}^\ell\gamma_5^{}\big)\ell
\;+\; {\rm H.c.}
\end{eqnarray}
can produce one of the best fits to the \,$b\to s$\, data if the Wilson coefficients
\,$C_i^\ell=C_i^{\rm SM}+C_i^{\ell,\rm NP}$\,
contain NP effects mainly in $C_{9,10}^{\mu,\rm NP}$ which satisfy the condition
\,$C_9^{\mu,\rm NP}=-C_{10}^{\mu,\rm NP}\sim-0.5$.\,
In this Lagrangian, $\alpha_{\rm e}^{}$ and $G_{\rm F}$ denote the usual fine structure
and Fermi constants, $V_{ts,tb}$ are elements of the Cabibbo-Kobayashi-Maskawa (CKM)
mixing matrix, \,$P_L^{}=(1-\gamma_5)/2$,\, and at the $m_b$ scale
\,$C_9^{\rm SM}\simeq-C_{10}^{\rm SM}\simeq4.2$\, universally for all charged leptons.
In contrast, dimension-six lepton-quark operators with tensor structures are excluded
by the measurements and (pseudo)scalar operators can explain them only with much
fine-tuning\,\,\cite{Hiller:2014yaa}, while the counterpart of ${\mathcal L}_{\rm SM+NP}$
with right-handed quark chirality leads to much poorer fits to
the data\,\,\cite{Altmannshofer:2014rta,Altmannshofer:2015sma}.

In view of the lepton nonuniversal nature of $C_i^{\ell,\rm NP}$ and the size of
$C_i^{\mu,\rm NP}$ relative to $C_i^{\rm SM}$, if the tentative indications of NP are
substantiated by upcoming experiments, one generally expects that there can be
\,$b\to s$\, transitions which violate lepton-flavor symmetry and have rates within
reach of searches in the near future\,\,\cite{Glashow:2014iga}.\footnote{\baselineskip=12pt%
The general arguments of \cite{Glashow:2014iga} regarding lepton-flavor violation do not hold for
certain models, such as those studied in \cite{Celis:2015ara}.}
Such a possibility for lepton flavor violation (LFV), and other LFV phenomena that could
have connections to the interactions of concern, have been examined further in
the literature in the contexts of various NP scenarios\,\,\cite{Buras:2014fpa,lfvmodels}.

In this paper, we also take these anomalies in \,$b\to s$\, data to be due to NP and explore
some of the potential consequences for a variety of rare meson decays with LFV.
To do so, we adopt the framework of so-called minimal flavor violation (MFV), which is based
on the hypothesis that Yukawa couplings are the only sources for the breaking of flavor and
$CP$ symmetries\,\,\cite{mfv1,D'Ambrosio:2002ex}, as flavor-dependent quark interactions
beyond the SM are empirically ruled out if they cause substantial flavor-changing
neutral currents.
Although the implementation of the MFV principle for quarks is straightforward, there is no
unique way to extend it to the lepton sector, as the SM alone does not accommodate LFV and
it is still unknown whether neutrinos are Dirac or Majorana particles.
Since there is now compelling evidence for neutrino masses and mixing~\cite{pdg}, it is
interesting to formulate leptonic MFV by incorporating ingredients beyond the SM that can
account for this observation~\cite{Cirigliano:2005ck}.
Thus, here we consider the SM expanded with the addition of three heavy right-handed neutrinos
as well as effective dimension-six quark-lepton operators with MFV built-in.
The heavy neutrinos participate in the usual seesaw mechanism to endow light neutrinos with
Majorana masses.
We will focus on a couple of such operators which can bring about the NP contributions
mentioned earlier and satisfy the MFV criterion in both the quark and
lepton sectors.\footnote{Various scenarios of leptonic MFV have been discussed in
the literature\,\,\cite{Branco:2006hz,mlfv,He:2014fva}.}
We will examine how these operators may contribute to a number of rare \,$b\to s$ and
\,$b\to d$ decays with LFV as well as the rare kaon decays \,$K_L\to e\mu$,\,
$K\to\pi e\mu$,\, and \,$K\to\pi\nu\nu$.\,

In the next section, after briefly reviewing the MFV framework, we introduce the NP operators
of interest.
In Section\,\,\ref{numerics} we write down the decay amplitudes and proceed with
our numerical analysis.
We provide our conclusions in Section$\;$\ref{conclusion}.
Additional information and lengthy formulas are relegated to an appendix.

\section{Operators with minimal flavor violaton\label{mfv}}

In the SM supplemented with three right-handed Majorana neutrinos, the renormalizable
Lagrangian for fermion masses can be written as
\begin{eqnarray} \label{Lm}
{\mathcal L}_{\rm m}^{} &\,=\,& -(Y_u)_{kl}^{}\,\overline{Q}_{k,L\,}^{}U_{l,R\,}^{} \tilde H
- (Y_d)_{kl}^{}\,\overline{Q}_{k,L\,}^{}D_{l,R\,}^{} H
- (Y_\nu)_{kl}^{}\,\overline{L}_{k,L\,}^{}\nu_{l,R\,}^{}\tilde H
- (Y_e)_{kl}^{}\,\overline{L}_{k,L\,}^{}E_{l,R\,}^{} H
\nonumber \\ && \!-~
\tfrac{1}{2}\, (M_\nu)_{kl}^{}\,\overline{\nu^{\rm c}_{k,R}}\,\nu_{l,R}^{}
\;+\; {\rm H.c.} ~,
\end{eqnarray}
where summation over \,$k,l=1,2,3$\, is implicit, $Y_{u,d,\nu,e}$ are matrices for
the Yukawa couplings, $Q_{k,L}$ $(L_{k,L})$ denote left-handed quark (lepton) doublets,
$U_{l,R}$ and $D_{l,R\,}$ $\bigl(\nu_{l,R}^{}$ and $E_{l,R}\bigr)$ represent right-handed
up- and down-type quarks (neutrinos and charged leptons), respectively, $H$\,\,stands for
the Higgs doublet, \,$\tilde H=i\tau_2^{}H^*$\, with $\tau_2^{}$ being the second Pauli matrix,
$M_\nu$~is a matrix for the Majorana masses of\,\,$\nu_{l,R}^{}$, and
\,$\nu^{\rm c}_{k,R}\equiv(\nu_{k,R})^{\rm c}$,\, the superscript referring to charge
conjugation.
With the nonzero elements of $M_\nu$ chosen to be much greater than those
of\,\,$v Y_\nu/\sqrt2$, the seesaw mechanism becomes operational\,\,\cite{seesaw1},
which leads to the light neutrinos' mass
matrix\,\,\,\mbox{$m_\nu^{}=-(v^2/2)\, Y_\nu^{}M_\nu^{-1}Y_\nu^{\rm T}=
U_{\scriptscriptstyle\rm PMNS\,}^{}\hat m_{\nu\,}^{}U_{\scriptscriptstyle\rm PMNS}^{\rm T}$},\,
where \,$v\simeq246$\,GeV\, is the Higgs's vacuum expectation value,
$U_{\scriptscriptstyle\rm PMNS}$ denotes the Pontecorvo-Maki-Nakagawa-Sakata
(PMNS~\cite{pmns}) matrix, and \,$\hat m_\nu^{}={\rm diag}\bigl(m_1^{},m_2^{},m_3^{}\bigr)$\,
contains the light neutrinos' eigenmasses $m_{1,2,3}^{}$.
This allows one to pick the form~\cite{Casas:2001sr}
\begin{eqnarray} \label{Ynum}
Y_\nu^{} \,\,=\,\,
\frac{i\sqrt2}{v}\,U_{\scriptscriptstyle\rm PMNS\,}^{}\hat m^{1/2}_\nu OM_\nu^{1/2}~,
\end{eqnarray}
where $O$ is in general a complex matrix satisfying \,$OO^{\rm T}=\openone$,\, the right-hand
side being a 3$\times$3 unit matrix.

We assume that the right-handed neutrinos are degenerate in mass,
\begin{eqnarray}
M_\nu^{} \,\,=\,\, {\mathcal M}\,{\rm diag}(1,1,1) \;.
\end{eqnarray}
The MFV hypothesis\,\,\cite{D'Ambrosio:2002ex,Cirigliano:2005ck} then implies that
${\mathcal L}_{\rm m}$ is formally invariant under the global flavor group
\,${\mathcal G}_f^{}=G_q^{}\times G_\ell^{}$,\, where
\,$G_q^{}={\rm SU}(3)_Q\times{\rm SU}(3)_U\times{\rm SU}(3)_D$\, and
\,$G_\ell={\rm SU}(3)_L\times{\rm O}(3)_\nu\times{\rm SU}(3)_E$.\,
This entails that $Q_{k,L}$, $U_{k,R}$, $D_{k,R}$, $L_{k,L}$, $\nu_{k,R}$, and $E_{k,R}$
belong to the fundamental representations of their respective flavor groups,
\begin{eqnarray}
Q_L^{} &\,\to\,& V_Q^{}Q_L^{} \,, ~~~~ ~~~ U_R^{} \,\to\, V_U^{}U_R^{} \,, ~~~~ ~~~
D_R^{} \,\to\, V_D^{}D_R^{} \,, ~~~~ ~~~ \nonumber \\
L_L^{} &\,\to\,&  V_L^{}L_L^{} \,, ~~~~~~~ \nu_R^{} \,\to\, {\mathcal O}_\nu^{}\nu_R^{} \,, ~~~~~~~
E_R^{} \,\to\, V_E^{}E_R^{} \,,
\end{eqnarray}
where \,$V_{Q,U,D,L,E}\in{\rm SU}(3)_{Q,U,D,L,E}$\, and \,${\mathcal O}_\nu\in{\rm O}(3)_\nu$\,
is an orthogonal real matrix\,\,\cite{D'Ambrosio:2002ex,Cirigliano:2005ck,Branco:2006hz}.
Furthermore, under ${\mathcal G}_f^{}$ the Yukawa couplings transform in the spurion sense
according to
\begin{eqnarray}
Y_u^{} \,\to\, V_Q^{}Y_u^{}V^\dagger_U \,, ~~~~~ Y_d^{} \,\to\, V_Q^{}Y_d^{}V^\dagger_D \,, ~~~~~
Y_\nu^{} \,\to\, V_L^{}Y_\nu^{}{\mathcal O}_\nu^{\rm T} \,, ~~~~~
Y_e^{} \,\to\, V_L^{}Y_e^{}V^\dagger_E \,.
\end{eqnarray}

Taking advantage of the symmetry under ${\mathcal G}_f^{}$, we can work in the basis where
\begin{eqnarray}
Y_d \,\,=\,\, \frac{\sqrt2}{v}\;{\rm diag}\bigl(m_d^{},m_s^{},m_b^{}\bigr) \,, ~~~~ ~~~
Y_e \,\,=\,\, \frac{\sqrt2}{v}\;{\rm diag}\bigl(m_e^{},m_\mu^{},m_\tau^{}\bigr)
\end{eqnarray}
and the fermion fields $U_k$, $D_k$, $\tilde\nu_{k,L}$, $\nu_{k,R}$, and $E_k$ refer to
the mass eigenstates.
More explicitly, \,$(U_1,U_2,U_3)=(u,c,t)$,\, $(D_1,D_2,D_3)=(d,s,b)$,\,
and \,$(E_1,E_2,E_3)=(e,\mu,\tau)$.\,
We can then express $Q_{k,L}$, $L_{k,L}$, and $Y_u$ in relation to the CKM matrix
$V_{\scriptscriptstyle\rm CKM}$ and $U_{\scriptscriptstyle\rm PMNS}$~as
\begin{eqnarray} \label{QL}
Q_{k,L}^{} = \left(\!\begin{array}{c}
\bigl(V^\dagger_{\scriptscriptstyle\rm CKM}\bigr)_{kl\,}U_{l,L}^{} \\
D_{k,L}^{} \end{array}\!\right) , ~~~~
L_{k,L}^{} = \left(\!\begin{array}{c} (U_{\scriptscriptstyle\rm PMNS})_{kl\,}^{}
\tilde\nu_{l,L}^{} \vspace{2pt} \\ E_{k,L}^{} \end{array}\!\right) , ~~~~
Y_u = \frac{\sqrt2}{v}\,V^\dagger_{\scriptscriptstyle\rm CKM}\,
{\rm diag}\bigl(m_u^{},m_c^{},m_t^{}\bigr) \,.
\end{eqnarray}

To put together effective Lagrangians beyond the SM with MFV built-in, one inserts products of
the Yukawa matrices among the relevant fields to construct ${\mathcal G}_f^{}$-invariant
operators that are singlet under the SM gauge group~\cite{D'Ambrosio:2002ex,Cirigliano:2005ck}.
Of interest here are the matrix products
\begin{eqnarray} \label{ABq}
\textsf{A}_q^{} &\,=\,& Y_u^{}Y_u^\dagger \,\,=\,\, V^\dagger_{\scriptscriptstyle\rm CKM}\,
{\rm diag}\bigl(y_u^2,y_c^2,y_t^2\bigr)\, V_{\scriptscriptstyle\rm CKM}^{} \,, \hspace{5em}
\textsf{B}_q^{} \,\,=\,\, Y_d^{}Y_d^\dagger \,\,=\,\, {\rm diag}\bigl(y_d^2,y_s^2,y_b^2\bigr) \,, \\
\textsf{A}_\ell^{} &\,=\,& Y_\nu^{}Y_\nu^\dagger \,\,=\,\, \frac{2\mathcal M}{v^2}\,
U_{\scriptscriptstyle\rm PMNS\,}^{} \hat m^{1/2}_\nu O O^\dagger\hat m^{1/2}_\nu
U_{\scriptscriptstyle\rm PMNS}^\dagger \,, ~~~~~~~~\,
\textsf{B}_\ell^{} \,\,=\,\, Y_e^{}Y_e^\dagger \,\,=\,\,
{\rm diag}\bigl(y_e^2,y_\mu^2,y_\tau^2\bigr) \,, ~~~~ ~~~ \label{ABl}
\end{eqnarray}
where \,$y_f^{}=\sqrt 2\,m_f^{}/v$.\,
Since the biggest eigenvalues of $\textsf{A}_q$ and $\textsf{B}_q$ are, respectively,
\,$y_t^2\sim1$\, and \,$y_b^2\sim3\times10^{-4}$\, at the $W$-boson
mass scale, for our purposes we can devise objects containing at most two powers of
$\textsf{A}_q$ and neglect contributions with $\textsf{B}_q$, as higher powers of
$\textsf{A}_q$ can be related to lower ones by means of the Cayley-Hamilton
identity\,\,\cite{Colangelo:2008qp}.
As for $\textsf{A}_\ell$, to maximize the NP effects we assume that the right-handed neutrinos'
mass $\mathcal M$ is sufficiently large to make the maximum eigenvalue of $\textsf{A}_\ell$
equal to unity, which fulfills the perturbativity requirement\,\,\cite{He:2014fva,Colangelo:2008qp}.
Hence, as in the quark sector, we will keep terms up to order $\textsf{A}_\ell^2$ and
drop those with $\textsf{B}_\ell$, whose elements are at most \,$y_\tau^2\sim1\times10^{-4}$\,.
Accordingly, the pertinent building blocks are
\begin{eqnarray} \label{Delta}
\Delta_q^{} \,\,=\,\, \zeta^{}_0\openone + \zeta^{}_{1\,}\textsf{A}_q^{}
+ \zeta^{}_{2\,}\textsf{A}_q^2 \,, ~~~~~~~
\Delta_\ell^{} \,\,=\,\, \xi^{}_0\openone + \xi^{}_{1\,}\textsf{A}_\ell^{}
+ \xi^{}_{2\,}\textsf{A}_\ell^2 \,,
\end{eqnarray}
where in our model-independent approach $\zeta_{0,1,2}^{}$ and $\xi_{0,1,2}^{}$ are free
parameters expected to be at most of ${\mathcal O}(1)$.
Hence one or more of them may be suppressed or vanish, depending on the underlying theory.
Since these parameters have negligible imaginary components\,\,\cite{He:2014fva,Colangelo:2008qp},
we can make the approximations
\,$\Delta_q^\dagger=\Delta_q^{}$\, and \,$\Delta_\ell^\dagger=\Delta_\ell^{}$.\,

It follows that the ${\mathcal G}_f^{}$-invariant dimension-six operators which are SM gauge
singlet and of the type that can readily give rise to the NP terms in Eq.\,(\ref{Lsmnp}) are
\begin{eqnarray}
{\mathcal O}_1^{} \,\,=\,\, \overline{Q}_L^{}\gamma_{\eta\,}^{}\Delta_{q1\,}^{}Q_L^{}\,
\overline{L}_L^{}\gamma^\eta\Delta_{\ell1\,}^{}L_L^{} \,, ~~~~~~~
{\mathcal O}_2^{} \,\,=\,\, \overline{Q}_L^{}\gamma_{\eta\,}^{}\Delta_{q2\,}^{}\tau_{a\,}^{}
Q_L^{}\,\overline{L}_L^{}\gamma^\eta\Delta_{\ell2\,}^{}\tau_{a\,}^{}L_L^{} \,,
\end{eqnarray}
where the objects $\Delta_{qj}$ and $\Delta_{\ell j}$ are, respectively, of the same form as
$\Delta_q$ and $\Delta_\ell$ in Eq.\,(\ref{Delta}), but have their own independent coefficients
$\zeta_{rj}^{}$ and $\xi_{rj}^{}$, and \,$a=1,2,3$\, is implicitly summed over.
These generalize operators that were previously introduced under the assumption of MFV only in
the quark\,\,\cite{D'Ambrosio:2002ex} or lepton\,\,\cite{Cirigliano:2005ck} part.
The MFV effective Lagrangian of interest is then
\begin{eqnarray} \label{Lmfv}
{\mathcal L}_{\rm MFV}^{} \,\,=\,\, \frac{1}{\Lambda^2}({\mathcal O}_1\,+\,{\mathcal O}_2) \;,
\end{eqnarray}
where the mass scale $\Lambda$ characterizes the heavy NP underlying these
interactions.
In general, one could also consider dimension-six (pseudo)scalar or tensor operators,
or operators with a right-handed quark current, that fulfill the MFV requirements and
contribute to \,$b\to s\ell\bar\ell'$\, transitions.
However, since as discussed in Section \ref{intro} the current data do not favor such types
of operators, their contributions can be neglected.

Before going into further details, it is instructive to go over the flavor structures
of ${\mathcal O}_1^{}$ and\,\,${\mathcal O}_2^{}$.
Expanding them in terms of the upper and lower components of the left-handed doublets
$Q_k^{}$ and $L_k^{}$ and suppressing the gamma matrices, we arrive at
\begin{eqnarray}
{\mathcal O}_1^{} &=&
\big(\hat\Delta_{q1}\big)_{mn}^{}(\Delta_{\ell1})_{kl}^{}
\big( \bar U_m^{}U_{n\,}^{}\bar\nu_k^{}\nu_l^{} + \bar U_m^{}U_{n\,}^{}\bar E_k^{}E_l^{} \big)
+ (\Delta_{q1})_{mn}^{}(\Delta_{\ell1})_{kl}^{} \big(\bar D_m^{}D_{n\,}^{}\bar\nu_k^{}\nu_l^{}
+ \bar D_m^{}D_{n\,}^{}\bar E_k^{}E_l^{} \big) \,,
\vphantom{|_{\int^|}} \nonumber \\
{\mathcal O}_2^{} &=&
\big(\hat\Delta_{q2}\big)_{mn}(\Delta_{\ell2})_{kl}^{} \big(
\bar U_m^{}U_{n\,}^{}\bar\nu_k^{}\nu_l^{}-\bar U_m^{}U_{n\,}^{}\bar E_k^{}E_l^{} \big)
- \bigl(\Delta_{q2}\bigr)_{mn}(\Delta_{\ell2})_{kl}^{} \big(
\bar D_m^{}D_{n\,}^{}\bar\nu_k^{}\nu_l^{}-\bar D_m^{}D_{n\,}^{}\bar E_k^{}E_l^{} \big)
\nonumber \\ && \! +\;
2 \big(V^\dagger_{\scriptscriptstyle\rm CKM}\hat\Delta_{q2}\bigr)_{mn}
(\Delta_{\ell2})_{kl}^{}\,\bar D_m^{}U_{n\,}^{}\bar\nu_k^{}E_l^{}
+ 2 \bigl(\hat\Delta_{q2}^{}V_{\scriptscriptstyle\rm CKM}^{}\bigr)_{mn}
(\Delta_{\ell2})_{kl}^{}\,\bar U_m^{}D_{n\,}^{}\bar E_k^{}\nu_l^{} \;, \label{OO'}
\end{eqnarray}
where
\,$\hat\Delta_{qj}^{}=
V_{\scriptscriptstyle\rm CKM\,}^{}\Delta_{qj}^{}V^\dagger_{\scriptscriptstyle\rm CKM}
=\zeta_{0j}^{}\openone + \zeta_{1j}^{}\,{\rm diag}\bigl(0,0,y_t^2\bigr)
+ \zeta_{2j}^{}\,{\rm diag}\bigl(0,0,y_t^4\bigr)$\,
with the $y_{u,c}^{}$ terms having been dropped, summation over \,$k,l,m,n=1,2,3$\, is
implicit, and \,$\nu_k^{}=(U_{\scriptscriptstyle\rm PMNS})_{kn\,}^{}\tilde\nu_n^{}$\,
represents a\,\,flavor eigenstate.
Hence, in the approximations we have made, ${\mathcal L}_{\rm MFV}$ does not cause
flavor-changing transitions between up-type quarks.
In contrast, flavor changes among down-type quarks can occur with either charged leptons or
neutrinos being emitted, but according to Eq.\,(\ref{OO'}) the two operators contribute
differently to the two types of processes, which will be treated in more detail below.
We will especially deal with the exclusive decays of
\,$b\to q\ell\bar\ell'$\, and \,$b\to q\nu\nu'$\, for \,$q=s,d$\,
and \,$\ell\neq\ell'$,\, as well as \,$s\to d\ell\bar\ell'$\, and \,$s\to d\nu\nu'$.\,

\section{Decay amplitudes and numerical analysis\label{numerics}}

In the presence of ${\mathcal L}_{\rm MFV}$, one can generalize Eq.\,(\ref{Lsmnp}) to
\begin{eqnarray} \label{b2qll'}
{\mathcal L}_{bq\ell\ell'}^{} \,\,=\,\,
\frac{\sqrt2\,\alpha_{\rm e\,}^{}\lambda_{qb\,}^{}G_{\rm F}^{}}{\pi}\, C_{\ell\ell'}^{}\,
\overline{q}_{\;\!}\gamma^\eta P_L^{}b\,\overline{\ell}_{\,\!}\gamma_\eta^{}P_L^{}\ell'
\;+\; {\rm H.c.} ~,
\end{eqnarray}
where \,$\lambda_{qb}^{}=V_{tq}^*V_{tb}^{}$\, is the CKM factor, \,$q=s,d$,\, and
\begin{eqnarray} \label{Cll'}
C_{\ell\ell'}^{} \,\,=\,\, \delta_{\ell\ell'\,}^{} C_9^{\rm SM}
\,+\, {\textsf c}_{\ell\ell'}^{} \,,
\end{eqnarray}
having set \,$C_{10}^{\rm SM}=-C_9^{\rm SM}$.\,
Accordingly, for \,$\ell\ell'=E_kE_l$\, one can derive from ${\mathcal L}_{\rm MFV}$
\begin{eqnarray} \label{cll'}
\textsf{c}_{E_kE_l}^{} \,\,=\,\, \tilde{\textsf{c}}_{E_kE_l}^{(1)} \,+\,
\tilde{\textsf{c}}_{E_kE_l}^{(2)}
\,\,=\,\, \tilde{\xi}_0^+\delta_{kl}^{} + \tilde{\xi}_1^+\big(\textsf{A}_\ell^{}\big)_{kl} +
\tilde{\xi}_2^+\big(\textsf{A}_\ell^2\big)_{kl} \,,
\end{eqnarray}
where $\tilde{\textsf{c}}_{E_kE_l}^{{\scriptscriptstyle(}j\scriptscriptstyle)}$ belongs to
${\mathcal O}_j$ and $\tilde{\xi}_{0,1,2}^+$ are given by
\begin{eqnarray} \label{tcll'}
\tilde{\xi}_r^+ \,\,=\,\, \tilde{\xi}_{r1}^{} + \tilde{\xi}_{r2}^{} \,, ~~~~~
r \,\,=\,\, 0,1,2 \,, ~~~~ ~~~
\tilde{\xi}_{rj}^{} \,=\,\, \frac{\pi_{\,}\big(\zeta_{1j\,}^{}y_t^2 +
\zeta_{2j\,}^{}y_t^4\big)\xi_{rj}^{}}
{\sqrt2\,\alpha_{\rm e}^{}\Lambda^{2\,}G_{\rm F}^{}} \,, ~~~~~
j \,\,=\,\, 1,2 \,,
\end{eqnarray}
the terms with $y_{u,c}^{}$ in $\tilde{\xi}_{rj}^{}$ having been dropped.
It follows that \,$|C_{\ell\ell'}^{}|=|C_{\ell'\ell}^{}|$.\,
Based on Eq.\,(\ref{b2qll'}), one can write down the corresponding Lagrangian
${\mathcal L}_{sd\ell\ell'}$ for \,$s\to d\ell^-\ell^{\prime+}$.\,

From the foregoing, we can derive the contributions of ${\mathcal L}_{\rm MFV}$ to the amplitudes
for a number of \,$b\to s\ell^-\ell^{\prime+}$\, transitions with \,$\ell\neq\ell'$.\,
Thus, with \,$C_{\ell\ell'}^{}={\textsf c}_{\ell\ell'}^{}$,\,
for \,$\bar B\to\bar K\ell^-\ell^{\prime+}$\, we obtain
\begin{eqnarray} \label{MB2Kll'}
{\mathcal M}_{\bar B\to\bar K\ell\bar\ell'}^{} \,\,=\,\,
\frac{-\alpha_{\rm e}^{}\lambda_{sb\,}^{}G_{\rm F\,}^{}{\textsf c}_{\ell\ell'}^{}}{\sqrt2\,\pi}
\Bigg[ \bigl(p_B^\eta+p_K^\eta\bigr)F_1+\frac{m_B^2-m_K^2}{\dot{s}}(F_0-F_1)\dot{p}^\eta
\Bigg] \overline{\ell}_{\,\!}\gamma_\eta^{}P_L^{}\ell' \,,
\end{eqnarray}
where $p_{B,K}^{}$ and \,$\dot{p}=p_\ell^{}+p_{\ell'}^{}$\, denote the four-momenta of the mesons
and dilepton, respectively, and $F_{0,1}$ stand for the form factors in the hadronic matrix
element $\bigl\langle\bar K\bigr|\bar s\gamma^\eta(1-\gamma_5)b\bigl|\bar B\bigr\rangle$,
which are described in Appendix\,\,\ref{rates} and depend on the Lorentz-invariant
\,$\dot{s}=\dot{p}^2$.\,
For \,$\bar B\to\bar K^*\ell^-\ell^{\prime+}$\, we arrive at
\begin{eqnarray} \label{MB2K*ll'}
{\mathcal M}_{\bar B\to\bar K^*\ell\bar\ell'}^{} &=&
\frac{-\alpha_{\rm e}^{}\lambda_{sb\,}^{}G_{\rm F\,}^{}{\textsf c}_{\ell\ell'}^{}}{\sqrt2\,\pi}
\Bigl[ {\mathbb A}\,\epsilon_{\eta\varsigma\chi\omega}^{}\,
\varepsilon^{*\varsigma}p_{B\,}^\chi p_{K^*}^\omega
- i{\mathbb C}\,\varepsilon_\eta^*
+ i{\mathbb D}\,\varepsilon^*\!\cdot\!\dot{p}\, \big(p_B^{}+p_{K^*}^{}\big)_\eta
\nonumber \\ && \hspace{7em} +\;
i{\mathbb H}\,\varepsilon^*\!\cdot\!\dot{p}\,\dot{p}_\eta^{} \Bigr]
\overline{\ell}_{\,\!}\gamma^\eta P_L^{}\ell' \,,
\vphantom{|_{\int_\int^|}^{}} \\
{\mathbb A} &=& 2_{\,\!}V/(m_B+m_{K^*}) \,, ~~~ ~~
{\mathbb C} \,=\, A_{1\,}(m_B+m_{K^*}) \,, ~~~ ~~
{\mathbb D} \,=\, A_2/(m_B+m_{K^*}) \,, \nonumber \\
\dot{s}_{\,}{\mathbb H} &\,=\,&
{\mathbb C}-{\mathbb D}\,\big(m_B^2-m_{K^*}^2\big)-2A_0^{}m_{K^*}^{} \,,
\end{eqnarray}
where $V$ and $A_{0,1,2}$ are the form factors for
$\bigl\langle\bar K^*\bigr|\bar s\gamma_\eta(1-\gamma_5)b\bigl|\bar B\bigr\rangle$,
which are defined in Appendix\,\,\ref{rates}.
We look at \,$\bar B_s\to\phi\ell^-\ell^{\prime+}$\, as well, which has an amplitude
${\mathcal M}_{\bar B_s\to\phi\ell\bar\ell'}$ analogous to
${\mathcal M}_{\bar B\to\bar K^*\ell\bar\ell'}$.
For \,$\bar B_s\to\ell^-\ell^{\prime+}$\, we find
\begin{eqnarray} \label{MB2ll'}
{\mathcal M}_{\bar B_s\to\ell\bar\ell'}^{} \,\,=\,\, \frac{i\alpha_{\rm e}^{}\lambda_{sb\,}^{}
f_{B_s}^{}G_{\rm F\,}^{}{\textsf c}_{\ell\ell'}^{}}{2\sqrt2\,\pi}\, \overline{\ell}
\big[m_{\ell'}^{}-m_\ell^{}+\big(m_{\ell'}^{}+m_\ell^{}\big)\gamma_5^{}\big]\ell' \,,
\end{eqnarray}
where $f_{B_s}^{}$ is the $B_s$ decay constant, which is also defined in Appendix\,\,\ref{rates}.

The MFV Lagrangian in Eq.\,(\ref{Lmfv}) generates lepton-flavor-violating
\,$b\to d\ell^-\ell^{\prime+}$\, processes with the same coefficient
${\textsf c}_{\ell\ell'}^{}$ in Eq.\,(\ref{Cll'}), but a different CKM factor, $\lambda_{db}$.
This makes it of interest to include them in our study, particularly
\,$\bar B\to\pi\ell^-\ell^{\prime+}$,\, $\bar B\to\rho\ell^-\ell^{\prime+}$,\, and
\,$\bar B_d\to\ell^-\ell^{\prime+}$.\,
Generally, there could be other dimension-six MFV operators that also contribute to
\,$b\to d\ell^-\ell^{\prime+}$\, and therefore can enhance or reduce the impact
of ${\textsf c}_{\ell\ell'}^{}$.
Hereafter, we focus on the possibility that the effects of such operators are unimportant.
Under these assumptions, the rates of these decay channels have expressions similar in form
to those for \,$\bar B\to K\ell^-\ell^{\prime+}$,\, $\bar B\to\bar K^*\ell^-\ell^{\prime+}$,\,
and\,\,\,$\bar B_s\to\ell^-\ell^{\prime+}$,\, respectively,
but are comparatively smaller because \,$|\lambda_{db}/\lambda_{sb}|^2\sim1/22$.\,

Before starting our numerical calculation, we need to specify our choices further.
Since the ${\textsf A}_\ell$ matrix in Eq.\,(\ref{ABl}) can be realized in many different
ways, we concentrate on the least complicated possibility that $O$ is a~real orthogonal
matrix, in which case
\begin{eqnarray} \label{Alchoice}
{\sf A}_\ell^{} \,\,=\,\, \frac{2_{\,}\mathcal M}{v^2}\,U_{\scriptscriptstyle\rm PMNS\,}^{}
\hat m_{\nu\,}^{}U_{\scriptscriptstyle\rm PMNS}^\dagger ~.
\end{eqnarray}
For $U_{\scriptscriptstyle\rm PMNS}$, we employ the standard parametrization\,\,\cite{pdg},
with its elements being determined from the results of a recent fit to global neutrino data
in Ref.\,\cite{Gonzalez-Garcia:2014bfa}, which depend on whether neutrino masses have a normal
hierarchy (NH), \,$m_1^{}<m_2^{}<m_3^{}$,\, or an inverted one (IH), \,$m_3^{}<m_1^{}<m_2^{}$.\,
Since the empirical information on the absolute scale of $m_{1,2,3}$ is still far from
precise~\cite{pdg}, for definiteness we pick \,$m_1=0$\, $(m_3=0)$\, in
the NH (IH) case.
Requiring the biggest eigenvalue of ${\sf A}_\ell$ to be unity then implies
\,${\mathcal M}\simeq6.1\times10^{14}$\,GeV.\,
For the elements of $V_{\scriptscriptstyle\rm CKM}$, we adopt the results of the latest
fit performed in Ref.\,\,\cite{ckmfit}.
With these numbers, we can evaluate the branching fractions of the decay modes discussed
earlier using the rate formulas collected in Appendix\,\,\ref{rates}, which also describes
our choices for the relevant hadronic form factors and decay constants.

We now attempt to attain the largest branching fractions of the \,$b\to q\ell\bar\ell'$\,
decays of interest, for \,$\ell\neq\ell'$,\, under our MFV framework by scanning the space
of the $\tilde{\xi}_r^+$ parameters, which enter the NP term,
${\textsf c}_{\ell\ell'}^{}$, in the Wilson coefficient $C_{\ell\ell'}^{}$ according to
Eqs.\,\,(\ref{Cll'})-(\ref{tcll'}).
This amounts to maximizing $|{\textsf c}_{\ell\ell'}^{}|$.
Simultaneously we need to impose the pertinent restrictions from the existing $b$-meson data.
Thus, based on the results of an analysis of the latest \,$b\to s$\, measurements including
the LHCb anomalies\,\,\cite{Altmannshofer:2014rta,Altmannshofer:2015sma}, we require
\begin{eqnarray} \label{C1}
{\textsf c}_{ee}^{} \,=\, 0 \;, ~~~~ ~~~
\mbox{$-0.71$} \,<\, {\textsf c}_{\mu\mu}^{} \,<\, -0.35 \;,
\end{eqnarray}
which can lead to one of the best fits to the data\,\,\cite{Altmannshofer:2015sma}.
Nevertheless, it is possible to let ${\textsf c}_{ee}^{}$ have some nonvanishing value
as well\,\,\cite{Altmannshofer:2014rta}.
Accordingly, we alternatively impose
\begin{eqnarray} \label{C2}
0 \,<\, {\textsf c}_{ee}^{} \,<\, 0.3 \;, ~~~~ ~~~
\mbox{$-0.65$} \,<\, {\textsf c}_{\mu\mu}^{} \,<\, -0.45 \;,
\end{eqnarray}
which is well within the allowed 1$\sigma$ best-fit region in the second plot of
Fig.\,6 in Ref.\,\cite{Altmannshofer:2014rta}.

In addition, since ${\textsf c}_{\ell\ell'}^{}$ for a specific pair of $\ell$ and
\,$\ell'\neq\ell$\, affects also \,$s\to d\ell\bar\ell'$\, processes, we need to take into
account the available experimental bounds, which may imply complementary limitations on
${\textsf c}_{\ell\ell'}^{}$.
Indeed, if like before other dimension-six MFV operators do not give rise to competitive
effects on
\,$s\to d\ell\bar\ell'$,\, we find in kaon data that the branching-fraction
limit\,\,\cite{pdg}
\,${\mathcal B}(K_L\to e^\pm\mu^\mp)_{\rm exp}^{}=\big({\mathcal B}(K_L\to e^+\mu^-)
+{\mathcal B}(K_L\to e^-\mu^+)\big)_{\rm exp}<4.7\times10^{-12}$\,
can translate into the strongest restriction on ${\textsf c}_{\ell\ell'}^{}$ among
lepton-flavor-violating meson decays.
Thus, applying Eq.\,(\ref{GKL2em}) in the appendix, with the central values of the input
parameters, we then extract
\begin{eqnarray} \label{Cem}
|{\textsf c}_{e\mu}|^2 \,<\, 0.16 \,,
\end{eqnarray}
which we will also impose.

\begin{table}[b]
\begin{tabular}{|c||c|c|} \hline
\multirow{3}{*}{~~ Decay mode ~~} & \multicolumn{2}{c|}{Branching fractions} \\ \cline{2-3} &
\multirow{2}{*}{\,$\begin{array}{c}\rm Measured\,\,upper\,\,limit\vspace{-3pt}\\
\rm at\;90\%\,CL~\mbox{\cite{pdg,hfag}}\end{array}$\,}
& \multirow{2}{*}{~ Prediction ~} \\ & & \\ \hline\hline
$B\to K e^\pm\mu^\mp\vphantom{|^{\int_|^|}}$   & $3.8\times10^{-8}$  & $9.7\times10^{-9}$ \\
$B\to K^*e^\pm\mu^\mp\vphantom{|^{\int_|^|}}$  & $5.1\times10^{-7}$  & $2.4\times10^{-8}$ \\
$B_s\to\phi e^\pm\mu^\mp\vphantom{|^{\int_|^|}}$            & --     & $2.4\times10^{-8}$ \\
$B_s\to e^\pm\mu^\mp\vphantom{|^{\int_|^|}}$   & $1.1\times10^{-8}$  &
$2.9\times10^{-11}\vphantom{|_{\int_|^0}}$
\\ \hline
$B\to\pi e^\pm\mu^\mp\vphantom{|^{\int_|^|}}$  & $9.2\times10^{-8}$  & $4.1\times10^{-10}$ \\
$B\to\rho e^\pm\mu^\mp\vphantom{|^{\int_|^|}}$ & $3.2\times10^{-6}$  & $1.1\times10^{-9}$ \\
$B^0\to e^\pm\mu^\mp\vphantom{|^{\int_|^|}}$   & $2.8\times10^{-9}$  &
$8.9\times10^{-13}\vphantom{|_{\int_|^0}}$
\\ \hline
$K^+\to\pi^+e^-\mu^+\vphantom{|^{\int_|^|}}$   & $1.3\times10^{-11}$ & $3.6\times10^{-14}$ \\
$K_L^{}\to\pi^0e^\pm\mu^\mp\vphantom{|^{\int_|^|}}$ & $7.6\times10^{-11}$ &
$4.5\times10^{-14}\vphantom{|_{\int_|^0}}$
\\ \hline
\end{tabular}
\caption{Predicted upper limits on the branching fractions of exclusive meson decays involving
$e\mu$ in the final states, calculated with $|{\textsf c}_{e\mu}|$ from the empirical
limit \,${\mathcal B}\big(K_L\to e^\pm\mu^\mp\big)_{\rm exp}<4.7\times10^{-12}$\,
at 90\% confidence level\,\,\cite{pdg}, under the assumption that the effects of
operators ${\cal O}_{1,2}$ dominate these processes.
For comparison, the experimental counterparts are also displayed if available.\label{b2qem}}
\end{table}

In our scans of the $\tilde{\xi}_r^+$ parameter space to maximize $|{\textsf c}_{\ell\ell'}|$,
we find that the bound in Eq.\,(\ref{Cem}) is always reached.
This implies that it can already be used to estimate the largest branching fractions of various
\,$\bar b\to\bar q e\mu$\, decays and \,$K\to\pi e\mu$\, within our MFV scenario with
${\cal O}_{1,2}$ taken to be the main operators responsible.
We display the results in the third column of Table\,\,\ref{b2qem} and compare them with
the corresponding experimental limits if available.\footnote{In conformity to the experimental
reports \cite{Edwards:2002kq}, the \,$B\to K^{(*)}e\mu$\, prediction in this table is
the simple average over the $B^+$ and $B^0$ channels, namely
\,${\cal B}\big(B\to K^{{\scriptscriptstyle(}*\scriptscriptstyle)}e^\pm\mu^\mp\big)=
\big({\cal B}(B^+\to K^{{\scriptscriptstyle(}*{\scriptscriptstyle)}+}e^\pm\mu^\mp)
+ {\cal B}(B^0\to K^{{\scriptscriptstyle(}*{\scriptscriptstyle)}0}e^\pm\mu^\mp)\big)/2$,\,
whereas the \,$B\to\pi e\mu$\, prediction is from
\,${\cal B}\big(B\to\pi e^\pm\mu^\mp\big)=\big({\cal B}(B^+\to\pi^+e^\pm\mu^\mp)
+ 2{\cal B}(B^0\to\pi^0e^\pm\mu^\mp)\big)/2$\,
and similarly for \,$B\to\rho e^\pm\mu^\mp$.}
One observes that the predicted \,${\cal B}(B\to K e^\pm\mu^\mp)$ is only 4 times
below its measured bound and, therefore, may be probed in near-future searches.
For the other modes, the predictions are lower than their experimental
counterparts by more than an order of magnitude.

For the decay channels with \,$\ell\ell'=e\tau$\, or \,$\mu\tau$,\, one can entertain
many different possibilities.
For several of them, we present the results listed in Table\,\,\ref{Cetmt},
where we have separated those obtained under the constraint in either
Eq.\,(\ref{C1}) or Eq.\,(\ref{C2}), besides the requirement in Eq.\,(\ref{Cem}).
Furthermore, to get the numbers without (within) parentheses in the table we have
employed the central values of the neutrino mixing parameters from
Ref.\,\cite{Gonzalez-Garcia:2014bfa} associated with the normal (inverted) hierarchy of
light neutrino masses, except for the Dirac $CP$-violation angle $\delta$ which has
a greater uncertainty than the other parameters.
To reflect this uncertainty, the left and right numbers inside the square brackets
have been computed with the minimum and maximum, respectively, of
\,$\delta/{\rm degree}=306^{+39}_{-70}\,\big(254^{+63}_{-62}\big)$\,
from Ref.\,\cite{Gonzalez-Garcia:2014bfa} in the NH (IH) case.

We remark that in Table\,\,\ref{Cetmt} each vertical set of results for
the $e\tau$ $(\mu\tau)$ channels comes from the same maximized value of
$|{\textsf c}_{e\tau}|$ $(|{\textsf c}_{\mu\tau}|)$.
However, for the $e\tau$ and $\mu\tau$ numbers in the same column, the respective sets
of $\tilde{\xi}_r^+$ values which maximize $|{\textsf c}_{e\tau}|$ and
$|{\textsf c}_{\mu\tau}|$ are not always the same.
Moreover, for the $e\tau$ $(\mu\tau)$ results in different columns, the $\tilde{\xi}_r^+$ sets
which maximize $|{\textsf c}_{e\tau}|$ $(|{\textsf c}_{\mu\tau}|)$ are generally also different,
as we are interested in attaining the biggest branching fractions in the different scenarios.
We note, in addition, that the results in the NH (IH) case follow from
\,$|{\textsf c}_{e\tau}|\mbox{\footnotesize$\,\lesssim\,$}0.9$\,\,(0.8)\, and
\,$|{\textsf c}_{\mu\tau}|\mbox{\footnotesize$\,\lesssim\,$}1.9$\,\,(1.4),\, with
\,$|\tilde{\xi}_0^+|\sim0.2$-0.7\,\,(1-2) and \,$|\tilde{\xi}_{1,2}^+|\sim4$-17\,\,(40-120),\,
implying that
\,$\Lambda/\big|\big(\zeta_{1j\,}^{}y_t^2+\zeta_{2j\,}^{}y_t^4\big)\xi_{rj}^{}\big|
\raisebox{1pt}{$^{1/2}$}>1.7$\,\,(0.65)\,\,TeV\,
if \,$\tilde{\xi}_{r1}^{}\sim\tilde{\xi}_{r2}^{}$.\,

\begin{table}[t]
\begin{tabular}{|c||c||c|c|} \hline
\multirow{3}{*}{$\begin{array}{c}\\ \rm Decay~mode\end{array}$} &
\multicolumn{3}{c|}{Branching fractions} \\ \cline{2-4} &
\multirow{2}{*}{$\begin{array}{c} \vspace{-17pt} \\
\rm Measured \vspace{-3pt}\\ \rm upper\,\,limit\vspace{-2pt}\\
\rm at~90\%\,CL~\mbox{\cite{hfag}}\end{array}$}
& \multicolumn{2}{c|}{Predictions$\vphantom{\int_{\int^|}^{\int_|^|}}$}
\\ \cline{3-4} & & (I) & (II)$\vphantom{\int_{\int^|}^{\int_|^|}}$ \\ \hline\hline\hline
$B^+\to K^+e^\pm\tau^\mp\vphantom{|^{\int_|^|}}$    & $3.0\times10^{-5}$ & \,
$[0.9,2.7]\;([1.5,2.2])\times10^{-8}$ \, & \, $[1.1,3.5]\;([2.0,2.6])\times10^{-8}$ \,
\\
$B^+\to K^{*+}e^\pm\tau^\mp\vphantom{|^{\int_|^|}}$ & --                 &
$[1.7,5.3]\;([3.0,4.3])\times10^{-8}$    &    $[2.2,6.9]\;([3.9,5.1])\times10^{-8}$
\\
$B_s\to\phi e^\pm\tau^\mp\vphantom{|^{\int_|^|}}$   & --                 &
$[1.7,5.0]\;([2.8,4.1])\times10^{-8}$    &    $[2.1,6.6]\;([3.7,4.8])\times10^{-8}$
\\
$B_s\to e^\pm\tau^\mp\vphantom{|^{\int_|^|}}$       & --$\vphantom{|_{\int_|^0}}$ &
$[0.9,2.6]\;([1.5,2.1])\times10^{-8}$    &    $[1.1,3.5]\;([1.9,2.5])\times10^{-8}$
\\ \hline
$B^+\to\pi^+e^-\tau^+\vphantom{|^{\int_|^|}}$       & $2.0\times10^{-5}$ &
$[0.2,0.7]\;([0.4,0.5])\times10^{-9}$    &    $[0.3,0.9]\;([0.5,0.6])\times10^{-9}$
\\
$B^+\to\rho^+e^\pm\tau^\mp\vphantom{|^{\int_|^|}}$      & -- &
$[0.8,2.4]\;([1.4,2.0])\times10^{-9}$    &    $[1.0,3.2]\;([1.8,2.3])\times10^{-9}$
\\
$B^0\to e^\pm\tau^\mp\vphantom{|^{\int_|^|}}$  & $2.8\times10^{-5}\vphantom{|_{\int_|^0}}$ &
$[0.3,0.8]\;([0.5,0.7])\times10^{-9}$    &    $[0.3,1.1]\;([0.6,0.8])\times10^{-9}$
\\ \hline\hline
$B^+\to K^+\mu^\pm\tau^\mp\vphantom{|^{\int_|^|}}$  & $4.8\times10^{-5}$ &
$[0.6,0.9]\;([0.4,0.5])\times10^{-7}$    &    $[0.8,1.4]\;([0.5,0.8])\times10^{-7}$
\\
$\;B^+\to K^{*+}\mu^\pm\tau^\mp\;\vphantom{|^{\int_|^|}}$ & --           &
$[1.1,1.8]\;([0.7,1.0])\times10^{-7}$    &    $[1.5,2.8]\;([1.0,1.5])\times10^{-7}$
\\
$B_s\to\phi\mu^\pm\tau^\mp\vphantom{|^{\int_|^|}}$\, & --           &
$[1.1,1.7]\;([0.7,1.0])\times10^{-7}$    &    $[1.5,2.6]\;([0.9,1.4])\times10^{-7}$
\\
$B_s\to\mu^\pm\tau^\mp\vphantom{|^{\int_|^|}}$       & --$\vphantom{|_{\int_|^0}}$ &
$[0.6,0.9]\;([0.4,0.5])\times10^{-7}$    &    $[0.8,1.4]\;([0.5,0.8])\times10^{-7}$
\\ \hline
$B^+\to\pi^+\mu^\pm\tau^\mp\vphantom{|^{\int_|^|}}$  & $7.2\times10^{-5}$ &
$[2.8,4.6]\;([1.8,2.6])\times10^{-9}$    &    $[3.9,6.9]\;([2.4,3.8])\times10^{-9}$
\\
$B^+\to\rho^+\mu^\pm\tau^\mp\vphantom{|^{\int_|^|}}$ & -- &
$[5.1,8.4]\;([3.2,4.8])\times10^{-9}$    &    $[7.1,13]\; ([4.4,7.0])\times10^{-9}$
\\
$B^0\to\mu^\pm\tau^\mp\vphantom{|^{\int_|^|}}$  & $2.2\times10^{-5}\vphantom{|_{\int_|^0}}$ &
$[1.7,2.8]\;([1.1,1.6])\times10^{-9}$    &    $[2.4,4.3]\;([1.5,2.4])\times10^{-9}$
\\ \hline
\end{tabular}
\caption{Predicted upper limits on the branching fractions of exclusive $b$-meson decays
involving  $(e,\mu)\tau$ in the final states, computed with the maximal
$|{\textsf c}_{(e,\mu)\tau}|$ determined under the imposed constraint set
\,(I) ${\textsf c}_{ee}^{}=0$, $-0.71<{\textsf c}_{\mu\mu}^{}<-0.35$, and
$|{\textsf c}_{e\mu}|<0.4$\, or
\,(II) $0<{\textsf c}_{ee}^{}<0.3$, $-0.65<{\textsf c}_{\mu\mu}^{}<-0.45$,
and\,\,$|{\textsf c}_{e\mu}|<0.4$,\, as discussed in the text, under the assumption that
the effects of operators ${\cal O}_{1,2}$ dominate these lepton-flavor-violating processes.
The numbers without (within) parentheses correspond to neutrino mixing parameters
belonging to the normal (inverted) hierarchy of neutrino masses, whereas the left and right
numbers inside square brackets reflect the minimum and maximum empirical values of the Dirac
phase $\delta$ in $U_{\scriptscriptstyle\rm PMNS}$.
For comparison, the data are also displayed if available.\label{Cetmt}}
\end{table}

It is evident from Table\,\,\ref{Cetmt} that for each decay mode the results in the different
cases are roughly of similar size and differ from each other within only factors of\,\,4 or less.
More interestingly, we notice that a few of the predicted branching fractions of the $\mu\tau$
channels can be as high as a few times $10^{-7}$.
Although they are still at least about two orders of magnitude below the existing empirical
limits, which are presently not many, upcoming experiments will expectedly offer ample
opportunities to look for these decays and improve the data situation.

Since the restraint in Eq.\,(\ref{Cem}) would lessen if there were other operators having
destructive interference with ${\textsf c}_{e\mu}$ in the $K_L$ decay amplitude,
their presence would bring about a different set of predictions for the processes listed
in Tables\,\,\ref{b2qem} and\,\,\ref{Cetmt}.
Changes in the predictions would also occur if other operators significantly modified
\,$b\to d\ell\bar\ell'$.\,
Thus, our scenario in which ${\cal O}_{1,2}$ dominate these LFV transitions will be tested when
one or more of them are discovered and the acquired data compared to the predictions.

It is worth commenting, in addition, that the parameter values which are responsible for
the predictions above also translate into reductions in the rates of the lepton-flavor-conserving
\,$b\to(s,d)\tau^+\tau^-$\, decays by up to a few tens percent with respect to their SM estimates.
This implies that future observations of these processes with good precision will serve as
important complementary tests on our NP scenario.

Now, as indicated in the preceding section, ${\mathcal L}_{\rm MFV}$ in Eq.\,(\ref{Lmfv}) also
contributes to transitions with neutrinos in the final states.
Specifically for \,$b\to q\nu\nu'$\, decays the effective Lagrangian is given by
\begin{eqnarray}
{\mathcal L}_{bq\nu\nu'}^{} \,\,=\,\,
\frac{\sqrt2\,\alpha_{\rm e\,}^{}\lambda_{qb\,}^{}G_{\rm F}^{}}{\pi}\,
C_{\nu\nu'\,}^{}\bar q\gamma^\eta P_L^{}b\,\bar\nu\gamma_\eta^{}P_L^{}\nu'
\;+\; {\rm H.c.} \,, ~~~~ ~~~
C_{\nu\nu'}^{} \,\,=\,\, \delta_{\nu\nu'\,}^{}C_L^{\rm SM}
+ \textsf{c}_{\nu\nu'}^{} \,,
\end{eqnarray}
where \,$C_L^{\rm SM}\simeq-6.4$\, is the SM prediction\,\,\cite{Buras:2014fpa} and
$\textsf{c}_{\nu\nu'}^{}$ arises from NP.
From ${\mathcal L}_{\rm MFV}$, one then gets for \,$\nu\nu'=\nu_k^{}\nu_l^{}$\,
\begin{eqnarray} \label{cnn'}
\textsf{c}_{\nu_k\nu_l}^{} \,=\,\, \tilde{\textsf{c}}_{E_kE_l}^{(1)} \,-\,
\tilde{\textsf{c}}_{E_kE_l}^{(2)}
\,\,=\,\, \tilde{\xi}_{0~}^-\delta_{kl}^{} + \tilde{\xi}_1^-\big(\textsf{A}_\ell^{}\big)_{kl} +
\tilde{\xi}_2^-\big(\textsf{A}_\ell^2\big)_{kl} \,,
\end{eqnarray}
where $\tilde{\textsf{c}}_{E_kE_l}^{{\scriptscriptstyle(}1,2\scriptscriptstyle)}$ also enter
$\textsf{c}_{E_kE_l}^{}$ in Eq.\,(\ref{cll'}), but with the opposite relative sign,
and \,$\tilde{\xi}_r^-=\tilde{\xi}_{r1}^{}-\tilde{\xi}_{r2}^{}$.\,
Therefore, $\textsf{c}_{\nu\nu'}^{}$ and $\textsf{c}_{\ell\ell'}^{}$ are generally
independent of each other\,\,\cite{Buras:2014fpa}.

Since only the contributions of $\textsf{c}_{\nu_k\nu_l}$ with \,$k=l=1,2,3$\, can interfere
with the SM contribution to \,$b\to q\nu\nu$,\, and since the neutrinos are not detected,
it is straightforward to derive the ratio of branching fractions
\begin{eqnarray} \label{rB2Knn}
r_{B\to K^{(*)}\nu\nu}^{} \,\,=\,\,
\frac{{\mathcal B}\raisebox{1pt}{$($}\bar B\to\bar K^{(*)}\nu\nu\raisebox{1.3pt}{$)$}}
{{\mathcal B}\raisebox{1pt}{$($}\bar B\to\bar K^{(*)}\nu\nu\raisebox{1.3pt}{$)$}_{\rm SM}}
\,\,=\,\, \frac{1}{3}\;
\raisebox{4pt}{\footnotesize$\displaystyle\sum_k$} \Bigg( \bigg| 1 +
\frac{\textsf{c}_{\nu_k\nu_k}^{}}{C_L^{\rm SM}} \bigg|
\raisebox{10pt}{$^2$} \,+\,
\raisebox{4pt}{\footnotesize$\displaystyle\sum_{l\neq k}$}\,\bigg|
\frac{\textsf{c}_{\nu_k\nu_l}^{}}{C_L^{\rm SM}}
\bigg|\raisebox{9pt}{$^2$} \Bigg) \,.
\end{eqnarray}
These channels are not yet observed, but there are experimental limits on their branching
fractions\,\,\cite{hfag}.
Here the relevant bound is \cite{Buras:2014fpa}
\,${\mathcal B}(\bar B\to\bar K\nu\nu)_{\rm exp}^{}<4.3\,
{\mathcal B}(\bar B\to\bar K\nu\nu)_{\rm SM}^{}$,\, and so we can impose
\begin{eqnarray} \label{rB2Knn<}
r_{B\to K^{(*)}\nu\nu}^{} \,\,<\,\, 4.3 \,.
\end{eqnarray}

Their counterparts in the kaon sector, \,$K^+\to\pi^+\nu\nu$\, and \,$K_L\to\pi^0\nu\nu$,\,
are similarly affected by\,\,$\textsf{c}_{\nu\nu'}^{}$.
In its presence, the effective Lagrangian for \,$sd\nu\nu'$\, interactions is given by
\begin{eqnarray}
{\mathcal L}_{ds\nu\nu'}^{} \,\,=\,\,
-\frac{\sqrt2\,\alpha_{\rm e}^{}\lambda_{t\,}^{}G_{\rm F}^{}}{\pi\,s_{\rm w}^2}\,
X_{\ell\ell'}^{}\,\bar s\gamma^\eta P_L^{}d\,\bar\nu\gamma_\eta^{}P_L^{}\nu
\;+\; {\rm H.c.} \,, ~~~~ ~~~
X_{\nu\nu'}^{} \,\,=\,\, \delta_{\nu\nu'\,}^{}X_{\rm SM}^{}
\,+\, {\textsf x}_{\nu\nu'}^{} \,, ~~~~~~~
\end{eqnarray}
where \,$\lambda_t=\lambda_{sd}$, \,$s_{\rm w}^2=\sin^2\!\theta_{\rm W}=0.231$,
the SM term\,\,\cite{k2pnn}
\,$X_{\rm SM}^{}=X+
|V_{us}|^4\,{\rm Re}\big(V_{cs}^*V_{cd}^{}\big)P_c^{}/\lambda_t^{}$\,
consisting of 2 contributions from top- and charm-loop diagrams, respectively, with
\,$X=-C_L^{\rm SM}s_{\rm w}^2$\, and
\,$|V_{us}|^4\,{\rm Re}\big(V_{cs}^*V_{cd}^{}\big)=\lambda^7/2-\lambda^5$\, in
the Wolfenstein parametrization,
and \,${\textsf x}_{\nu\nu'}^{}=-{\textsf c}_{\nu\nu'}^{}s_{\rm w}^2$\, due to NP.
The latest predictions of the SM are\,\,\cite{Buras:2015qea}
\,${\mathcal B}(K^+\to\pi^+\nu\nu)_{\rm SM}=(9.11\pm0.72)\times10^{-11}$\, and
\,${\mathcal B}(K_L\to\pi^0\nu\nu)_{\rm SM}=(3.0\pm0.3)\times10^{-11}$.\,
Generalizing the expressions for these branching fractions to include
the ${\mathcal L}_{\rm MFV}$ effects, which modify the $X$ part, we obtain
\begin{eqnarray}
{\mathcal B}\raisebox{1pt}{$($}K^+\to\pi^+\nu\nu\raisebox{1pt}{$)$} &\,=\,&
\frac{\kappa_+'}{3_{\,\!}\lambda^{10}}\,\raisebox{3pt}{\footnotesize$\displaystyle\sum_k$}
\bigg\{ \big[{\rm Re}\lambda_t\,\big(X-\textsf{c}_{\nu_k\nu_k}s_{\rm w}^2\big)
+ \big(\lambda^7/2-\lambda^5\big)P_c \big]^2
\nonumber \\ && \hspace{9ex} +\;
({\rm Im}\lambda_t)^2\big(X-\textsf{c}_{\nu_k\nu_k}s_{\rm w}^2\big)^2 +
|\lambda_t|^2s_{\rm w}^4\, \raisebox{3pt}{\footnotesize$\displaystyle\sum_{l\neq k}$}\,
\big|\textsf{c}_{\nu_k\nu_l}^{}\big|^2 \bigg\} \,, ~~~~~~~
\vphantom{|_{\int_\int}^{}} \\ \label{BKL2pnn}
{\mathcal B}\raisebox{1pt}{$($}K_L\to\pi^0\nu\nu\raisebox{1pt}{$)$} &\,=\,&
\frac{\kappa_L^{}\,({\rm Im}\lambda_t)^2}{3_{\,\!}\lambda^{10}}\,
\raisebox{3pt}{\footnotesize$\displaystyle\sum_k$} \bigg[
\big(X-\textsf{c}_{\nu_k\nu_k}s_{\rm w}^2\big)^2 + s_{\rm w}^4\,
\raisebox{3pt}{\footnotesize$\displaystyle\sum_{l\neq k}$}\,
\big|\textsf{c}_{\nu_k\nu_l}^{}\big|^2 \bigg] \,,
\end{eqnarray}
where \,$\kappa_+'=0.997\times5.173\times10^{-11}$,\, $X=1.481$,\, $P_c=0.404$,\,
and \,$\kappa_L^{}=2.231\times10^{-10}$\, are the central values from Ref.\,\cite{Buras:2015qea}.
In view of the data\,\,\cite{Artamonov:2008qb}
\,${\mathcal B}(K^+\to\pi^+\nu\nu)_{\rm exp}=\big(17.3^{+11.5}_{-10.5}\big)\times10^{-11}$
and\,\,\cite{Ahn:2009gb}
\,${\mathcal B}(K_L\to\pi^0\nu\nu)_{\rm exp}<2.6\times10^{-8}$,\,
we may then demand only
\begin{eqnarray} \label{rK2pnn<}
0.7 \,\,<\,\, r_{K^+\to\pi^+\nu\nu}^{} \,\,=\,\,
\frac{{\mathcal B}\raisebox{1pt}{$($}K^+\to\pi^+\nu\nu\raisebox{1.3pt}{$)$}}
{{\mathcal B}\raisebox{1pt}{$($}K^+\to\pi^+\nu\nu\raisebox{1.3pt}{$)$}_{\rm SM}}
\,\,<\,\, 3.2 \,.
\end{eqnarray}
In addition, comparing Eqs.\,\,(\ref{rB2Knn}) and\,\,(\ref{BKL2pnn}), one infers
that $r_{B\to K^{(*)}\nu\nu}$ is equal to its \,$K_L\to\pi^0\nu\nu$\, counterpart,
$r_{K_L\to\pi^0\nu\nu}$.

For the parameter values that yield the examples in Table\,\,\ref{Cetmt},
if \,${\textsf c}_{\ell\ell'}^{\scriptscriptstyle(2)}=0$\, in Eqs.\,\,(\ref{cll'})
and\,\,(\ref{cnn'}), we find that \,$r_{B\to K^{(*)}\nu\nu}=r_{K_L\to\pi^0\nu\nu}$\, can be
as large as 1.22 (1.15) in the NH (IH) case,
whereas $r_{K^+\to\pi^+\nu\nu}$ can reach \,1.15\,\,(1.11).\,
If \,${\textsf c}_{\ell\ell'}^{\scriptscriptstyle(1)}=0$\, instead, the results above for
the decay modes with charged leptons are unchanged, but now the branching fractions of the modes
with neutrinos tend to be reduced by up to\,\,14\% with respect to their SM expectations.
All of these numbers are well within their corresponding restrictions in
Eqs.\,\,(\ref{rB2Knn<}) and\,\,(\ref{rK2pnn<}).

If both ${\textsf c}_{\ell\ell'}^{\scriptscriptstyle(1)}$ and
${\textsf c}_{\ell\ell'}^{\scriptscriptstyle(2)}$ are nonzero, then in general
$\textsf{c}_{\ell\ell'}^{}$ and $\textsf{c}_{\nu\nu'}^{}$ are not connected and,
consequently, they can be maximized independently.
In that case, our results in Table\,\,\ref{Cetmt} are still the same, but for the channels with
neutrinos, after scanning the $\tilde{\xi}_r^-$ parameter space subject to
Eqs.\,\,(\ref{rB2Knn<}) and\,\,(\ref{rK2pnn<}), we obtain a maximum value of
\,$r_{B\to K^{(*)}\nu\nu}=r_{K_L\to\pi^0\nu\nu}$\, that saturates its limit of\,\,4.3 and
$r_{K^+\to\pi^+\nu\nu}$ that reaches \,{\footnotesize$\sim$\,}3.1,\,
with \,$\tilde{\xi}_0^-\sim-7$\, and \,$|\tilde{\xi}_{1,2}^-|\sim0$.\,
Thus future measurements can offer significant checks on these predictions.

Finally, we mention that, in the approximations we made, ${\mathcal L}_{\rm MFV}$ produces
vanishing effects on the SM-dominated transitions
\,$b\to(u,c)\nu\ell$, \,$s\to u\nu\ell$,\, and \,$c\to(d,s)\nu\bar\ell$.\,
Although its contribution to \,$t\to b\nu\bar\ell$\, is nonzero,
\begin{eqnarray}
{\cal M}_{t\to b\nu\bar\ell}^{\rm MFV} \,\,=\,\,
\frac{-2 V_{tb}^* \big(\zeta_{02}^{}+\zeta_{12\,}^{}y_t^2+\zeta_{22\,}^{}y_t^4\big)
(\Delta_{\ell2})_{kl}^{}}{\Lambda^2}\,
\bar b\gamma^\eta P_L^{}t\,\bar\nu_k^{}\gamma_\eta^{}P_L^{}E_l^{} \,,
\end{eqnarray}
we estimate that its branching fraction is only under $10^{-3}$ for optimistic choices of
the parameter values, and so it is much smaller than
\,${\cal B}(t\to b W^+\to b\nu\ell^+)_{\rm SM}^{}\sim0.1$\,
which is consistent with the data\,\,\cite{pdg}.

\section{Conclusions\label{conclusion}}

We have entertained the possibility that the anomalies recently detected in the measurements
of rare \,$b\to s$\, processes are NP signals and explored some of the potential implications
for a\,\,number of exclusive lepton-flavor-violating meson decays.
Adopting the effective theory framework of MFV based on the SM plus 3 heavy right-handed
neutrinos participating in the seesaw mechanism, we concentrate on a couple of dimension-six
4-fermion operators which accommodate the interactions that can yield one of the best fits to
the \,$b\to s$\, data.
Assuming that these operators conform to the MFV hypothesis in both their quark and lepton
parts, we evaluate their effects on various meson decays that violate lepton-flavor
symmetry, subject to additional relevant $b$-meson and kaon constraints.
For simplification, we further assume that other dimension-six MFV operators do not induce
competitive contributions to these transitions.
This scenario can be tested when they are discovered and the resulting predictions confronted
with the acquired data.

With the preceding premises, our numerical work shows that for decays with charged leptons
in the final states ${\sf c}_{e\mu}$ is the most restricted among the Wilson coefficients
${\sf c}_{\ell\ell'}$ for \,$\ell\neq\ell'$,\, its strictest bound being supplied by
the experimental limit for \,$K_L\to e\mu$.\,
Interestingly, among the other modes considered, the resulting prediction for
\,${\cal B}(B\to K e\mu)$ is the closest to its empirical bound, being only 4 times smaller,
and therefore may be probed by forthcoming searches.
Moreover, the predicted branching fractions of the exclusive \,$b\to s\mu\tau$\,
decays can be as large as a few times\,\,$10^{-7}$ and hence may be examined as well in
near-future experiments.
For decay channels with neutrinos in the final states, \,$B\to K^{(*)}\nu\nu$\, and
\,$K\to\pi\nu\nu$,\, we find that they are comparatively far less restricted and that
the impact of the MFV operators can enhance their branching fractions by up to a few times
with respect to the SM expectations.
Thus, planned measurements on these processes involving neutrinos will provide important
checks on the enhancements.

\acknowledgments

We would like to thank Xiao-Gang He for comments.
This work was supported in part by the MOE Academic Excellence Program (Grant No. 102R891505).

\appendix

\section{Decay rates\label{rates}}

In what follows we assume that the charged leptons $\ell$ and $\ell'$ in the final states
of the decays are different in flavor.
Furthermore, the decays are induced by left-handed current-current operators of the type
in Eq.\,(\ref{b2qll'}).

The hadronic matrix elements pertinent to the decay \,$\bar B\to\bar K\ell\bar\ell'$\, are then
\begin{eqnarray} \label{B->K}
\bigl\langle\bar K\bigr|\bar s\gamma^\eta b\bigl|\bar B\bigr\rangle \,\,=\,\,
\frac{m_B^2-m_K^2}{\hat s}\, \hat p^\eta\, F_0^{} +
\bigg(p_B^\eta+p_K^\eta-\frac{m_B^2-m_K^2}{\hat s}\,\hat p^\eta\bigg) F_1^{}
\end{eqnarray}
and \,$\langle\bar K|\bar s\gamma^\eta\gamma_5^{}b|\bar B\rangle=0$,\, where
$m_B^{}$\,\,$(m_K^{})$ and $p_B$\,\,$(p_K)$ are the $\bar B\;\bigl(\bar K\bigr)$ mass
and four-momentum, respectively, \,$\hat p=p_B^{}-p_K^{}$,\, and the form factors $F_{0,1}^{}$
depend on the Lorentz-invariant \,$\hat s=\hat p^2$\,  according to\,\,\cite{F0F1}
\begin{eqnarray}
F_0^{} &=& a_{00}^{}+a_{01}^{}z+a_{02}^{}z^2+a_{03}^{}z^3 \,, ~~~~ ~~~
F_1^{} \,=\,
\frac{a_{10}^{}+a_{11}^{}z+a_{12}^{}z^2-\frac{1}{3}\big(a_{11}^{}-2a_{12}^{}\big)z^3}
{1-\hat s/\big(m_B^{}+0.04578{\rm\;GeV}\big)\raisebox{1pt}{$^2$}} \,,
\nonumber \\
z &=& \frac{\sqrt{t_+-\hat s}-\sqrt{t_+-t_0}}{\sqrt{t_+-\hat s}+\sqrt{t_+-t_0}} \,, ~~~~~
t_{\pm} \,=\, \big(m_{B^+}^{}\pm m_{K^+}^{}\big)\raisebox{1pt}{$^2$} \,, ~~~~~
t_0 \,=\, \Big(1-\sqrt{1-t_-/t_+}\Big)t_+ \,, ~~~~~~~
\end{eqnarray}
the $a$'s being constants.
Numerically, we adopt
\,$\big(a_{00}^{},a_{01}^{},a_{02}^{},a_{03}^{}\big)=(0.54,-1.91,1.83,-0.02)$\, and
\,$\big(a_{10}^{},a_{11}^{},a_{12}^{}\big)=(0.43,-0.67,-1.12)$,\, which are their
central values from
Ref.\,\,\cite{Altmannshofer:2014rta}.
For the \,$\bar B\to\pi$\, form-factors, which are relevant to
\,$\bar B\to\pi\ell\bar\ell'$\, and defined analogously to $F_{0,1}$ in\,\,Eq.\,(\ref{B->K}),
we make use of the parametrization choice preferred in Ref.\,\cite{Ali:2013zfa}
as well as the relation
\,$\langle\pi^-|\bar d\gamma^\eta b|B^-\rangle
=\sqrt2\,\langle\pi^0|\bar d\gamma^\eta b|\bar B^0\rangle$\,
based on isospin symmetry.

Similarly, for \,$\bar B\to\bar K^*\ell\bar\ell'$\, the requisite matrix elements are
\begin{eqnarray} \label{B->K*}
\bigl\langle\bar K^*\bigr|\bar s\gamma_\eta^{}b\bigl|\bar B\bigr\rangle &=&
\frac{2 V^{}}{m_B^{}+m_{K^*}^{}}\, \epsilon_{\eta\rho\sigma\tau}^{}\,
\varepsilon^{*\rho} p_B^\sigma\, p_{K^*}^\tau \,,
\nonumber \\
\bigl\langle\bar K^*\bigr|\bar s\gamma^\eta\gamma_5^{}b\bigl|\bar B\bigr\rangle  &=&
2 i A_{0\,}^{}m_{K^*}^{}\, \frac{\varepsilon^*\cdot\tilde p}{\tilde s}\,\tilde p^\eta
+ i A_1^{}\bigl( m_B^{}+m_{K^*}^{} \bigr) \biggl(
\varepsilon^{*\eta}-\frac{\varepsilon^*\cdot\tilde p}{\tilde s}\,\tilde p^\eta \biggr)
\nonumber \\ && \! -\;
\frac{i A_2^{}\,\varepsilon^*\cdot\tilde p}{m_B^{}+m_{K^*}^{}}
\biggl( p_B^\eta + p_{K^*}^\eta - \frac{m_B^2-m_{K^*}^2}{\tilde s}\,\tilde p^\eta \biggr) \,,
\end{eqnarray}
where \,$\tilde p=p_B^{}-p_{K^*}^{}$\, and the form factors $V^{}$ and $A_{0,1,2}$ are
functions of \,$\tilde s=\tilde p^2$.\,
For \,$\bar B_s\to\phi\ell\bar\ell'$,\, the hadronic matrix elements have expressions similar
to those for \,$\bar B\to\bar K^*\ell\bar\ell'$.\,
In numerical work, we adopt the \,$\bar B\to\bar K^*$\, and \,$\bar B_s\to\phi$\, form-factors
available in Ref.\,\cite{Straub:2015ica} from combined fits to light-cone-sum-rule and lattice
results.
For the \,$b\to d$\, transition \,$\bar B\to\rho\ell\bar\ell'$,\, the form factors are defined
analogously to those in Eq.\,(\ref{B->K*}), we again utilize the parametrization
provided in\,\,Ref.\,\cite{Straub:2015ica}, and as in the \,$\bar B\to\pi$\, case we have
\,$\langle\rho^-|\bar d\gamma^\eta(1-\gamma_5)b|B^-\rangle
=\sqrt2\,\langle\rho^0|\bar d\gamma^\eta(1-\gamma_5)b|\bar B^0\rangle$.\,

The formulas in the last two paragraphs lead us to the amplitudes in
Eqs.\,\,(\ref{MB2Kll'}) and\,\,(\ref{MB2K*ll'}), respectively.
Subsequently, we arrive at the differential rates
\begin{eqnarray} \label{dGB2Kll'/ds}
\frac{d\Gamma_{\bar B\to\bar K\ell\bar\ell'}^{}}{d\dot{s}} &=&
\frac{\big|\alpha_{\rm e\,}^{}\lambda_{qb\,}^{}{\textsf c}_{\ell\ell'\,}^{}G_{\rm F}^{}
\big|\raisebox{1pt}{$^2$}\sqrt{f_1^{~}\hat g}}{1536_{\,}\pi^{5\,}m_{B\,}^3\dot{s}^3}
\Big[ F_1^2 f_2^{}\, \hat g \,+\,
F_0^2 f_3^{}\, \big(m_B^2-m_K^2\big)\raisebox{2pt}{$^{\!2}$} \Big] \,, ~~~~~~~
\\ \nonumber \\ \label{dGB2K*ll'/ds}
\frac{d\Gamma_{\bar B\to\bar K^*\ell\bar\ell'}^{}}{d\dot{s}} &=&
\frac{\big|\alpha_{\rm e\,}^{}\lambda_{qb\,}^{}{\textsf c}_{\ell\ell'\,}^{}G_{\rm F}^{}\big|^2
\sqrt{f_1^{~}\tilde g}}{6144_{\,}\pi^{5\,}m_{B\,}^3m_{K^*\,}^2\dot{s}^3} \Big\{
2\big[ \big({\mathbb A}^2\tilde g s+6_{\,}{\mathbb C}^2s
+ {\mathbb{CD}}_{\,\!}\tilde g\big) f_2^{} + 2 A_0^{2\,}f_3^{~}\tilde g \big] m_{K^*}^2
\nonumber \\ && \hspace{22ex} +\;
\big[ {\mathbb C}^2+2_{\,}{\mathbb{CD}}_{\,}\big(s-m_B^2\big)+{\mathbb D}^2\tilde g \big]
f_2^{~}\tilde g \Big\} \,, ~~~~
\end{eqnarray}
where \,$\dot{s}=(p_\ell^{}+p_{\ell'}^{})^2$,\,
\begin{eqnarray}
f_1^{} &=& {\mathcal K}\big(\dot{s},m_\ell^2,m_{\ell'}^2\big) \,, \hspace{9ex}
f_2^{} \,=\, 2\dot{s}^2 - \big(m_\ell^2+m_{\ell'}^2\big)\dot{s}
- \big(m_\ell^2-m_{\ell'}^2\big)\raisebox{1pt}{$^2$} \,,
\nonumber \\
\hat g &=& {\mathcal K}\big(m_B^2,m_K^2,\dot{s}\big) \,, \hspace{8.3ex}
f_3^{} \,=\, 3\big(m_\ell^2+m_{\ell'}^2\big) \dot{s}
- 3\big(m_\ell^2-m_{\ell'}^2\big)\raisebox{1pt}{$^2$} \,,
\nonumber \\
\tilde g &=& {\mathcal K}\big(m_B^2,m_{K^*}^2,\dot{s}\big) \,, \hspace{7.7ex}
{\mathcal K}(x,y,z) \,=\, x^2+y^2+z^2-2(x y+y z+x z) \,.
\end{eqnarray}
In computing the branching fractions, we integrate Eqs.\,\,(\ref{dGB2Kll'/ds}) and
(\ref{dGB2K*ll'/ds}) over the whole kinematical ranges
\,$(m_\ell+m_{\ell'})^2\le\dot{s}\le(m_B-m_K)^2$\,
and \,$(m_\ell+m_{\ell'})^2\le\dot{s}\le(m_B-m_{K^*})^2$,\, respectively.
We do likewise for \,$\bar B_s\to\phi\ell\bar\ell'$\, and
\,$\bar B\to(\pi,\rho)\ell\bar\ell'$.\,

To examine the purely leptonic decay \,$\bar B_q\to\ell\bar\ell'$,\, the hadronic matrix
elements we need are
\,$\big\langle0\big|\bar q\gamma^\eta b\big|\bar B_q\big\rangle=0$\, and
\,$\big\langle0\big|\bar q\gamma^\eta\gamma_5^{}b\big|\bar B_q\big\rangle=-i f_{B_q}^{}p_B^\eta$,\,
where $f_{B_q}^{}$ is the decay constant.
The amplitude in Eq.\,(\ref{MB2ll'}) then follows, leading to the decay rate
\begin{eqnarray} \label{GB2ll'}
\Gamma_{\bar B_q\to\ell\bar\ell'}^{} \,\,=\,\, \frac{\big|\alpha_{\rm e\,}^{}\lambda_{qb\,}^{}
{\textsf c}_{\ell\ell'}^{~~\;}f_{B_q}^{}G_{\rm F}^{}\big|\raisebox{1pt}{$^2$\,}
{\mathcal K}^{1/2}\bigl(m_B^2,m_\ell^2,m_{\ell'}^2\bigr)}{32_{\,}\pi^{3\,}m_B^{}} \Bigg[
m_\ell^2+m_{\ell'}^2-\frac{\big(m_\ell^2-m_{\ell'}^2\big)\raisebox{1pt}{$^2$}}{m_B^2} \Bigg] \,.
\end{eqnarray}

The kaon reaction \,$K_L\to e^-\mu^+$\, proceeds from the components of
\,$K_L\simeq\big(K^0+\bar K^0\big)/\sqrt2$\, both decaying into \,$e^-\mu^+$.\,
Analogously to \,$B_q\to\ell\bar\ell'$,\, the necessary hadronic matrix elements are
\,$\big\langle0\big|\bar s\gamma^\eta d\big|K^0\big\rangle=
\big\langle0\big|\bar d\gamma^\eta s\big|\bar K^0\big\rangle=0$\, and
\,$\big\langle0\big|\bar s\gamma^\eta\gamma_5^{}d\big|K^0\big\rangle=
\big\langle0\big|\bar d\gamma^\eta\gamma_5^{}s\big|\bar K^0\big\rangle=-i f_K^{}p_K^\eta$,\,
where $f_K^{}$ is the kaon decay constant.
Neglecting $m_e^{}$, with the aid of Eq.\,(\ref{GB2ll'}) we then find
\begin{eqnarray} \label{GKL2em}
\Gamma_{K_L\to e\bar\mu}^{} \,\,=\,\, \frac{\big|\alpha_{\rm e\,}^{}{\textsf c}_{e\mu}^{~~\;}
f_{K\,}^{}G_{\rm F}^{}\,{\rm Re}_{\,\!}\lambda_{sd}^{}\big|\raisebox{1pt}{$^2$\,}
m_{K^0\,}^{}m_\mu^2}{16_{\,\!}\pi^3} \Bigg(1-\frac{m_\mu^2}{m_{K^0}^2}\Bigg)^{\!\!2} \,.
\end{eqnarray}

The amplitude for \,$K^+\to\pi^+e^-\mu^+$,\, also containing ${\textsf c}_{e\mu}$, is similar
to that for \,$K^+\to\pi^0\nu\mu^+$\, which arises mainly from SM interactions described
by \,${\cal L}_{su\nu\ell}^{\rm SM}=-\sqrt8\,G_{\rm F}^{}V_{us}^*\,
\bar s\gamma^\eta P_L^{}u\,\bar\nu\gamma_\eta^{}P_L^{}\mu\,+\,\rm H.c.$\,
It follows that one can conveniently express the branching fraction
${\cal B}(K^+\to\pi^+e^-\mu^+)$ in relation to the well-measured
\,${\cal B}(K^+\to\pi^0\nu\mu^+)_{\rm exp}=(3.353\pm0.034)\times10^{-2}$,\, upon assuming
isospin symmetry and neglecting\,\,$m_e^{}$, without having to know the \,$K\to\pi$\,
form-factors in great detail.
Thus, since
\,$\langle\pi^+|\bar s\gamma^\eta d|K^+\rangle
=\sqrt2\,\langle\pi^0|\bar s\gamma^\eta u|K^+\rangle$,\,
we arrive at
\begin{eqnarray}
{\cal B}(K^+\to\pi^+e^-\mu^+) \,\,\simeq\,\, \frac{|\alpha_{\rm e\,}^{}\lambda_{sd\,}^{}
{\textsf c}_{e\mu}^{}|^2}{2_{\,\!}\pi^{2\,}|V_{us}|^2}\,{\cal B}(K^+\to\pi^0\nu\mu^+)_{\rm exp}^{} \,.
\end{eqnarray}
Similarly, for \,$K_L\to\pi^0e^-\mu^+$,\, since
\,$\langle\pi^0|\bar d\gamma^\eta s|\bar K^0\rangle=-\langle\pi^0|\bar s\gamma^\eta d|K^0\rangle
=\langle\pi^0|\bar s\gamma^\eta u|K^+\rangle$,\,
we get
\begin{eqnarray}
{\cal B}(K_L\to\pi^0e^\pm\mu^\mp) \,=\, 2_{\,\!}{\cal B}(K_L\to\pi^0e^-\mu^+) \,\simeq\,
\frac{\tau_{K_L^{}\,}^{}|\alpha_{\rm e\,}^{}{\textsf c}_{e\mu}^{}\,{\rm Im}_{\,\!}\lambda_{sd}^{}|^2}
{\pi^{2\,}\tau_{K^+\,}^{}|V_{us}|^2}\,{\cal B}(K^+\to\pi^0\nu\mu^+)_{\rm exp}^{} \,. ~~~~~
\end{eqnarray}

In numerical applications of these rate formulas, we employ the $B^+$ and $B_{s,d}$
$\big(K_L$ and $K^+\big)$ lifetimes from Ref.\,\,\cite{hfag}\,\,(\cite{pdg}),
\,$\alpha_{\rm e}^{}=1/133$,\, and \,$G_{\rm F}=1.1664\times10^{-5}\rm\;GeV^2$.\,
For the decay constants, we adopt the central values of \,$f_{B_d}^{}=(190.5\pm4.2)$\,MeV,
\,$f_{B_s}^{}=(227.7\pm4.5)$\,MeV,\, and \,$f_K^{}=(156.3\pm0.9)$\,MeV\, from Ref.\,\cite{fBdfBs}.


\begin{thebibliography}{0}

\bibitem{Aaij:2014ora}
  R.~Aaij {\it et al.}  [LHCb Collaboration],
  Phys.\ Rev.\ Lett.\  {\bf 113}, 151601 (2014)  [arXiv:1406.6482 [hep-ex]].

\bibitem{Aaij:2013qta}
  R.~Aaij {\it et al.}  [LHCb Collaboration],
  Phys.\ Rev.\ Lett.\  {\bf 111}, 191801 (2013)  [arXiv:1308.1707 [hep-ex]].

\bibitem{LHCb:2015dla}
  The LHCb Collaboration,
Report No. LHCb-CONF-2015-002, March 2015.

\bibitem{Aaij:2013aln}
  R.~Aaij {\it et al.}  [LHCb Collaboration],
  JHEP {\bf 1307}, 084 (2013)  [arXiv:1305.2168 [hep-ex]];
  JHEP {\bf 1406}, 133 (2014)  [arXiv:1403.8044 [hep-ex]].

\bibitem{Descotes-Genon:2013wba}
  S.~Descotes-Genon, J.~Matias, and J.~Virto,
  Phys.\ Rev.\ D {\bf 88}, 074002 (2013)  [arXiv:1307.5683 [hep-ph]];
  W.~Altmannshofer and D.M.~Straub,
  Eur.\ Phys.\ J.\ C {\bf 73}, 2646 (2013)  [arXiv:1308.1501 [hep-ph]];
  F.~Beaujean, C.~Bobeth, and D.~van Dyk,
  Eur.\ Phys.\ J.\ C {\bf 74}, 2897 (2014); 3179(E) (2014)  [arXiv:1310.2478 [hep-ph]];
  T.~Hurth and F.~Mahmoudi,
  JHEP {\bf 1404}, 097 (2014)  [arXiv:1312.5267 [hep-ph]];
  R.~Alonso, B.~Grinstein, and J.~Martin Camalich,
  Phys.\ Rev.\ Lett.\  {\bf 113}, 241802 (2014)  [arXiv:1407.7044 [hep-ph]];
  D.~Ghosh, M.~Nardecchia, and S.A.~Renner,
  JHEP {\bf 1412}, 131 (2014)  [arXiv:1408.4097 [hep-ph]];
  T.~Hurth, F.~Mahmoudi, and S.~Neshatpour,
  JHEP {\bf 1412}, 053 (2014)  [arXiv:1410.4545 [hep-ph]];
  S.~Descotes-Genon, L.~Hofer, J.~Matias, and J.~Virto,
  arXiv:1503.03328 [hep-ph].

\bibitem{Hiller:2014yaa}
  G.~Hiller and M.~Schmaltz,
  Phys.\ Rev.\ D {\bf 90}, 054014 (2014)  [arXiv:1408.1627 [hep-ph]].

\bibitem{Altmannshofer:2014rta}
  W.~Altmannshofer and D.M.~Straub,
  Eur.\ Phys.\ J.\ C {\bf 75}, no. 8, 382 (2015)  [arXiv:1411.3161 [hep-ph]].

\bibitem{Altmannshofer:2015sma}
  W.~Altmannshofer and D.M.~Straub,
  arXiv:1503.06199 [hep-ph].

\bibitem{rsmodel}
Some of these issues and their implications have also been addressed exclusively
within specific models, such as in
  P.~Biancofiore, P.~Colangelo, and F.~De Fazio,
  Phys.\ Rev.\ D {\bf 89}, no. 9, 095018 (2014)  [arXiv:1403.2944 [hep-ph]];
  P.~Biancofiore, P.~Colangelo, F.~De Fazio, and E.~Scrimieri,
  Eur.\ Phys.\ J.\ C {\bf 75}, no. 3, 134 (2015)  [arXiv:1408.5614 [hep-ph]].

\bibitem{Lyon}
  J.~Lyon and R.~Zwicky,
  arXiv:1406.0566 [hep-ph].

\bibitem{Glashow:2014iga}
  S.L.~Glashow, D.~Guadagnoli, and K.~Lane,
  Phys.\ Rev.\ Lett.\  {\bf 114}, 091801 (2015)  [arXiv:1411.0565 [hep-ph]].

\bibitem{Celis:2015ara}
  A.~Celis, J.~Fuentes-Martin, M.~Jung, and H.~Serodio,
  arXiv:1505.03079 [hep-ph].

\bibitem{Buras:2014fpa}
  A.J.~Buras, J.~Girrbach-Noe, C.~Niehoff, and D.M.~Straub,
  JHEP {\bf 1502}, 184 (2015)  [arXiv:1409.4557 [hep-ph]].

\bibitem{lfvmodels}
  B.~Gripaios, M.~Nardecchia, and S.A.~Renner,
  JHEP {\bf 1505}, 006 (2015)  [arXiv:1412.1791 [hep-ph]];
  B.~Bhattacharya, A.~Datta, D.~London and S.~Shivashankara,
  Phys.\ Lett.\ B {\bf 742}, 370 (2015)  [arXiv:1412.7164 [hep-ph]];
  A.~Crivellin, G.~D'Ambrosio, and J.~Heeck,
  Phys.\ Rev.\ Lett.\  {\bf 114}, 151801 (2015)  [arXiv:1501.00993 [hep-ph]];
  Phys.\ Rev.\ D {\bf 91}, no. 7, 075006 (2015)  [arXiv:1503.03477 [hep-ph]];
  S.~Sahoo and R.~Mohanta,
  Phys.\ Rev.\ D {\bf 91}, no. 9, 094019 (2015)  [arXiv:1501.05193 [hep-ph]];
  I.~de Medeiros Varzielas and G.~Hiller,
  arXiv:1503.01084 [hep-ph];
  S.D.~Aristizabal, F.~Staub, and A.~Vicente,
  arXiv:1503.06077 [hep-ph];
  S.M.~Boucenna, J.W.F.~Valle, and A.~Vicente,
  arXiv:1503.07099 [hep-ph];
  D.~Becirevic, S.~Fajfer, and N.~Ko$\check{\rm s}$nik,
  arXiv:1503.09024 [hep-ph];
  A.~Crivellin, L.~Hofer, J.~Matias, U.~Nierste, S.~Pokorski, and J.~Rosiek,
  arXiv:1504.07928 [hep-ph].

\bibitem{mfv1}
  R.S.~Chivukula and H.~Georgi,
  Phys.\ Lett.\ B {\bf 188}, 99 (1987);
L.J.~Hall and L.~Randall,
  Phys.\ Rev.\ Lett.\  {\bf 65}, 2939 (1990);
  A.J.~Buras, P.~Gambino, M.~Gorbahn, S.~Jager, and L.~Silvestrini,
  Phys.\ Lett.\ B {\bf 500}, 161 (2001)  [hep-ph/0007085];
  A.J.~Buras,
  Acta Phys.\ Polon.\ B {\bf 34}, 5615 (2003)  [hep-ph/0310208];
  A.L.~Kagan, G.~Perez, T.~Volansky, and J.~Zupan,
  Phys.\ Rev.\ D {\bf 80}, 076002 (2009)  [arXiv:0903.1794 [hep-ph]].

\bibitem{D'Ambrosio:2002ex}
  G.~D'Ambrosio, G.F.~Giudice, G.~Isidori, and A.~Strumia,
  Nucl.\ Phys.\ B {\bf 645}, 155 (2002)  [hep-ph/0207036].

\bibitem{pdg}
  K.A.~Olive {\it et al.}  [Particle Data Group Collaboration],
  Chin.\ Phys.\ C {\bf 38}, 090001 (2014).

\bibitem{Cirigliano:2005ck}
  V.~Cirigliano, B.~Grinstein, G.~Isidori, and M.B.~Wise,
  Nucl.\ Phys.\ B {\bf 728}, 121 (2005)  [hep-ph/0507001].

\bibitem{Branco:2006hz}
G.C.~Branco, A.J.~Buras, S.~Jager, S.~Uhlig, and A.~Weiler,
  JHEP {\bf 0709}, 004 (2007)  [hep-ph/0609067];

\bibitem{mlfv}
  S.~Davidson and F.~Palorini,
  Phys.\ Lett.\ B {\bf 642}, 72 (2006)  [hep-ph/0607329];
M.B.~Gavela, T.~Hambye, D.~Hernandez, and P.~Hernandez,
  JHEP {\bf 0909}, 038 (2009)  [arXiv:0906.1461 [hep-ph]];
  A.S.~Joshipura, K.M.~Patel, and S.K.~Vempati,
  Phys.\ Lett.\ B {\bf 690}, 289 (2010)  [arXiv:0911.5618 [hep-ph]];
  R.~Alonso, G.~Isidori, L.~Merlo, L.A.~Munoz, and E.~Nardi,
  JHEP {\bf 1106}, 037 (2011)  [arXiv:1103.5461 [hep-ph]];
  D.~Aristizabal Sierra, A.~Degee, and J.F.~Kamenik,
  JHEP {\bf 1207}, 135 (2012)  [arXiv:1205.5547 [hep-ph]];
  X.G.~He, C.J.~Lee, J.~Tandean, and Y.J.~Zheng,
  Phys.\ Rev.\ D {\bf 91}, no. 7, 076008 (2015)  [arXiv:1411.6612 [hep-ph]].

\bibitem{He:2014fva}
  X.G.~He, C.J.~Lee, S.F.~Li, and J.~Tandean,
  Phys.\ Rev.\ D {\bf 89}, 091901 (2014)  [arXiv:1401.2615 [hep-ph]];
  JHEP {\bf 1408}, 019 (2014)  [arXiv:1404.4436 [hep-ph]].

\bibitem{seesaw1}
  P.~Minkowski,
  Phys.\ Lett.\  B {\bf 67}, 421 (1977);
T.~Yanagida, in {\it Proceedings of the Workshop on the Unified Theory and the Baryon Number in
the Universe}, edited by O.~Sawada and A.~Sugamoto (KEK, Tsukuba, 1979), p.~95;
  Prog.\ Theor.\ Phys.\  {\bf 64}, 1103 (1980);
M.~Gell-Mann, P.~Ramond, and R.~Slansky,
in {\it Supergravity}, edited by P.~van Nieuwenhuizen and D.~Freedman
(North-Holland, Amsterdam, 1979), p.~315;
  P.~Ramond,
  arXiv:hep-ph/9809459;
S.L.~Glashow, in {\it Proceedings of the 1979 Cargese Summer Institute on Quarks and Leptons},
edited by M.~Levy {\it et al}. (Plenum Press, New York, 1980), p.~687;
  R.N.~Mohapatra and G.~Senjanovic,
  Phys.\ Rev.\ Lett.\  {\bf 44}, 912 (1980);
J.~Schechter and  J.W.F.~Valle,
  Phys.\  Rev.\ D {\bf 22}, 2227 (1980);
  Phys.\  Rev.\ D {\bf 25}, 774 (1982).

\bibitem{pmns}
  B.~Pontecorvo,
Sov.\ Phys.\ JETP {\bf 26} (1968) 984
  [Zh.\ Eksp.\ Teor.\ Fiz.\  {\bf 53} (1968) 1717];
  Z.~Maki, M.~Nakagawa, and S.~Sakata,
  Prog.\ Theor.\ Phys.\  {\bf 28}, 870 (1962).

\bibitem{Casas:2001sr}
  J.A.~Casas and A.~Ibarra,
  Nucl.\ Phys.\ B {\bf 618}, 171 (2001)   [hep-ph/0103065].

\bibitem{Colangelo:2008qp}
  G.~Colangelo, E.~Nikolidakis, and C.~Smith,
  Eur.\ Phys.\ J.\ C {\bf 59}, 75 (2009)  [arXiv:0807.0801 [hep-ph]];
  L.~Mercolli and C.~Smith,
  Nucl.\ Phys.\ B {\bf 817}, 1 (2009)  [arXiv:0902.1949 [hep-ph]].

\bibitem{Gonzalez-Garcia:2014bfa}
  M.C.~Gonzalez-Garcia, M.~Maltoni, and T.~Schwetz,
  JHEP {\bf 1411}, 052 (2014)  [arXiv:1409.5439 [hep-ph]].

\bibitem{ckmfit}
  J.~Charles, O.~Deschamps, S.~Descotes-Genon, H.~Lacker, A.~Menzel, S.~Monteil, V.~Niess,
and J.~Ocariz {\it et al.},
  Phys.\ Rev.\ D {\bf 91}, no. 7, 073007 (2015)  [arXiv:1501.05013 [hep-ph]].
Online updates available at http://ckmfitter.in2p3.fr.

\bibitem{Edwards:2002kq}
  K.~W.~Edwards {\it et al.}  [CLEO Collaboration],
  Phys.\ Rev.\ D {\bf 65}, 111102 (2002)  [hep-ex/0204017];
B.~Aubert {\it et al.}  [BaBar Collaboration],
  Phys.\ Rev.\ D {\bf 73}, 092001 (2006)  [hep-ex/0604007];
  Phys.\ Rev.\ Lett.\  {\bf 99}, 051801 (2007)  [hep-ex/0703018].

\bibitem{hfag}
  Y.~Amhis {\it et al.}  [Heavy Flavor Averaging Group (HFAG) Collaboration],
  arXiv:1412.7515 [hep-ex];
Online updates available at http://www.slac.stanford.edu/xorg/hfag.

\bibitem{k2pnn}
  G.~Buchalla, A.J.~Buras, and M.E.~Lautenbacher,
  Rev.\ Mod.\ Phys.\  {\bf 68}, 1125 (1996)  [arXiv:hep-ph/9512380].
  F.~Mescia and C.~Smith,
  Phys.\ Rev.\ D {\bf 76}, 034017 (2007)  [arXiv:0705.2025 [hep-ph]].

\bibitem{Buras:2015qea}
  A.J.~Buras, D.~Buttazzo, J.~Girrbach-Noe, and R.~Knegjens,
  arXiv:1503.02693 [hep-ph].

\bibitem{Artamonov:2008qb}
  A.V.~Artamonov {\it et al.}  [E949 Collaboration],
  Phys.\ Rev.\ Lett.\  {\bf 101}, 191802 (2008)  [arXiv:0808.2459 [hep-ex]].

\bibitem{Ahn:2009gb}
  J.K.~Ahn {\it et al.}  [E391a Collaboration],
  Phys.\ Rev.\ D {\bf 81}, 072004 (2010)  [arXiv:0911.4789 [hep-ex]].

\bibitem{F0F1}
  C.~Bouchard {\it et al.}  [HPQCD Collaboration],
  Phys.\ Rev.\ D {\bf 88}, no. 5, 054509 (2013); {\bf 88}, no. 7, 079901(E) (2013)
[arXiv:1306.2384 [hep-lat]].

\bibitem{Ali:2013zfa}
  A.~Ali, A.Y.~Parkhomenko, and A.V.~Rusov,
  Phys.\ Rev.\ D {\bf 89}, no. 9, 094021 (2014)  [arXiv:1312.2523 [hep-ph]].

\bibitem{Straub:2015ica}
  A.~Bharucha, D.M.~Straub, and R.~Zwicky,
  arXiv:1503.05534 [hep-ph].

\bibitem{fBdfBs}
  S.~Aoki, Y.~Aoki, C.~Bernard, T.~Blum, G.~Colangelo,
  {\it et al.},  
  Eur.\ Phys.\ J.\ C {\bf 74}, 2890 (2014)  [arXiv:1310.8555 [hep-lat]].


\end{thebibliography}
\end{document}